\pdfoutput=1
\documentclass[showpacs,twocolumn, superscriptaddress,amsmath,amssymb,showkeys,prl,notitlepage]{revtex4-1} 
\usepackage[T1]{fontenc}
\usepackage[english]{babel}
\usepackage{graphicx}
\usepackage{amsmath,amssymb,mathrsfs}
\usepackage{hyperref}
\usepackage{caption}
\usepackage{mypackages}
\usepackage{mymacros}
\usepackage{todonotes}
\usepackage{lineno}
\newenvironment{ncabstract}{%
\begin{quote} \bf}
{\end{quote}}
\usetikzlibrary{matrix}
\definecolor{alain}{RGB}{240,226,182}

\definecolor{jordan}{RGB}{192,225,215}

\definecolor{philippe}{RGB}{90,150,175}

\definecolor{eric}{RGB}{250,160,130}

\begin{document}
\widetext
\title{Structurally Triggered Metal-Insulator Transition in Rare-Earth Nickelates}

\author{Alain Mercy} \affiliation{Theoretical Materials Physics, Q-MAT, CESAM, University of Li\`ege, B-4000 Li\`ege, Belgium}
\author{Jordan Bieder} \affiliation{Theoretical Materials Physics, Q-MAT, CESAM, University of Li\`ege, B-4000 Li\`ege, Belgium} \affiliation{CEA DAM-DIF, F-91297 Arpajon, France}
\author{Jorge \'I\~niguez} \affiliation{Department of Materials Research and Technology, Luxembourg Institute of Science and Technology, L-4362 Esch/Alzette, Luxembourg}
\author{Philippe Ghosez} \affiliation{Theoretical Materials Physics, Q-MAT, CESAM, University of Li\`ege, B-4000 Li\`ege, Belgium}

\vskip 0.25cm

%

\date{\today}



\maketitle

\setlength{\plotwidth}{\linewidth}

\begin{ncabstract}
Rare-earth nickelates form an intriguing series of correlated perovskite oxides. Apart from LaNiO$_\mathbf{3}$, they exhibit on cooling a sharp metal-insulator electronic phase transition, a concurrent structural phase transition and a magnetic phase transition toward an unusual antiferromagnetic spin order. Appealing for various applications, full exploitation of these compounds is still hampered by the lack of global understanding of the interplay between their electronic, structural and magnetic properties.  Here, we show from first-principles calculations that the metal-insulator transition of nickelates arises from the softening of an oxygen breathing distortion, structurally triggered by oxygen-octahedra rotation motions. The origin of such a rare triggered mechanism is traced back in their electronic and magnetic properties, providing a united picture. We further develop a Landau model accounting for the evolution of the metal-insulator transition in terms of the $\boldsymbol{{R}}$ cations and rationalising how to tune this transition by acting on oxygen rotation motions.\end{ncabstract}

First synthetized in 1971~\cite{Demazeau1971}, rare-earth nickelates ($R$NiO$_3$, with $R =$ rare earth) are appealing for various  applications~\cite{Shi2013,Zhou2016} and the possibility to tune their properties in epitaxial films and heterostructures~\cite{Middey2016} has recently fuelled an even larger interest~\cite{Forst2015,Kim2016,Grisolia2016}. 
$R$NiO$_3$ compounds belong to the family of perovskite oxides with a reference $Pm\bar{3}m$ cubic structure (Fig. 1a), nevertheless not directly observed.  Apart for LaNiO$_3$, which always keeps a metallic $R\bar{3}c$ phase and will not be further discussed here, all $R$NiO$_3$ adopt at reasonably high temperature a metallic $Pbnm$ orthorhombic phase~\cite{Catalan2008}. This phase, rather ubiquitous~\cite{Benedek2013} amongst perovskites with small Goldschmidt tolerance factor~\cite{Catalan2008}, 
$t$, corresponds to a distortion of the cubic structure arising from the appearance of (i) combined anti-phase rotations of the oxygen octahedra 
along the $x$ and $y$ directions, $R_{xy}$ (Fig. 1b), with the same amplitude $Q_{R}$  and (ii) in-phase rotations of the oxygen octahedra along 
$z$, $M_z$ (Fig. 1c), with amplitude $Q_{M}$. 

\begin{figure}
\begin{center}
\resizebox{7.5cm}{!}{\includegraphics{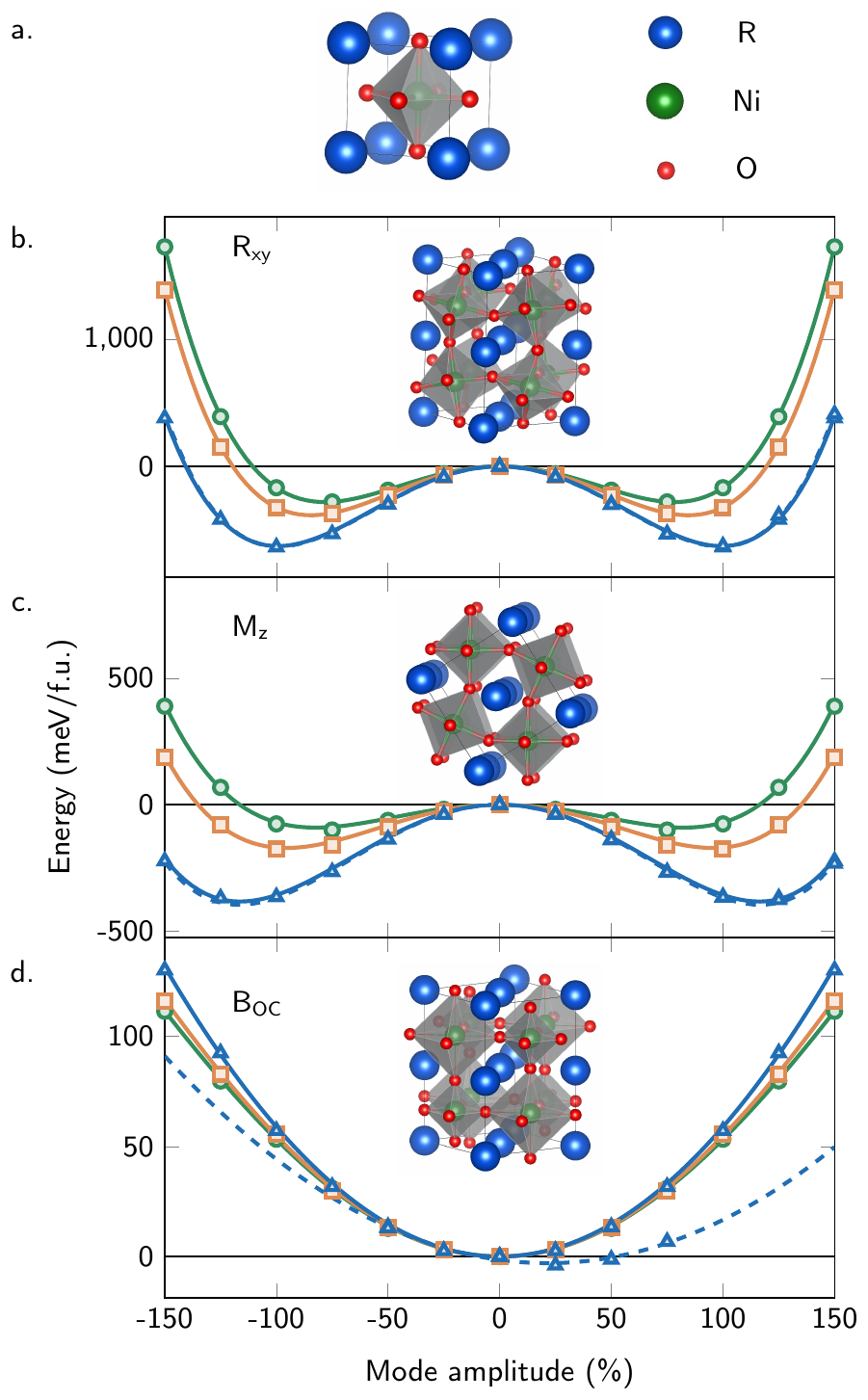}}
\end{center}
\caption{ {\bf Reference cubic perovskite structure and most relevant atomic distortions.} ({\bf a}) Sketch of the reference $Pm\bar{3}m$ cubic perovskite cell of $R$NiO$_3$ compound with $R$ at the corner,  Ni at the centre and O atoms at the middle of the faces, forming corner-shared octahedra surrounding the B atoms. Starting from this reference, three main atomic distortions drive the system successively to the $Pbnm$ and then $P2_1/n$ phases : ({\bf b}) anti-phase rotations of oxygen octahedra of same amplitudes about $x$ and $y$ axis ($R_{xy}$), ({\bf c}) in-phase rotations of oxygen octahedra about $z$ axis ($M_{z}$), ({\bf d}) breathing of the oxygen octahedra ($B_{OC}$). The energy wells associated to the freezing of individual distortion of increasing amplitude in the cubic phase are shown for different $R$ cations, associated to distinct tolerance factor $t$, and either a ferromagnetic (full line) or E'-type antiferromagnetic (dashed line, only for $R=$Y) spin arrangement :  YNiO$_3$ ($t =0.920$,  blue), GdNiO$_3$ ($t =0.938$,  orange) and SmNiO$_3$ ($t =0.947$,  green). The atomic distortions are normalised to their amplitude in the  $P2_1/n$ AFM-E' ground state of YNiO$_3$. Calculations are done for each compound in a cubic phase that has the same volume as the ground state.}
 \label{fig-1}
\end{figure}

In this phase, all Ni atoms are equivalent and formally in a Jahn-Teller active $d^7$ (likely $t_{2g}^6e_g^1$ low spin) configuration. Surprisingly, 
although compatible with the $Pbnm$ symmetry, cooperative Jahn-Teller distortions do not appear. Instead, at a temperature $T_{MI}$ which 
strongly evolves with the $R$ cation (i.e with $t$)~\cite{Medarde1997}, $R$NiO$_3$ compounds exhibit an electronic metal-insulator transition 
(MIT), which was shown to be concurrent with a structural transition from $Pbnm$ to monoclinic $P2_1/n$ symmetry \cite{Alonso1999}.  This symmetry lowering is accompanied 
with the appearance of a breathing distortion of the oxygen octahedra, $B_{OC}$ (Fig. 1d), which alternatively expand and contract with amplitude $Q_B$, according to a rock-salt pattern. This gives rise to two types of Ni sites with long and short Ni--O bonds respectively.  

At low temperature (100-200K), $R$NiO$_3$ compounds finally show a magnetic phase transition toward an unusual E'-type antiferromagnetic (AFM) spin 
order identified by a Bragg vector $\mathbf{q}=(1/4,1/4,1/4)$ in pseudocubic notation~\cite{Garcia-munoz1994,Alonso1999,Lee2011}. For large cations ($R =$ Nd and Pr),  the Neel 
temperature $T_N =  T_{MI}$ and the system goes directly from paramagnetic metal (PM-M) to  AFM insulator (AFM-I). For smaller $R$ cations, $T_N$ is much lower than $T_{MI}$; 
the two phase transitions are decoupled and the system goes through an intermediate paramagnetic insulating phase (PM-I). 

The origin of the MIT has been widely debated in the literature \cite{Torrance1992, Mizokawa2000,Raebiger2008,Park2012,Ruppen2015,Varignon2016}. It was sometimes interpreted as a  
charge disproportionation at Ni sites~\cite{Mazin2007}~: $2 (d^7) \rightarrow d^8+d^6$. However, the importance of Ni 3d -- O 2p hybridization -- i.e., transfer of electrons from O to Ni and formation of oxygen holes (\underline{L}) keeping Ni in a $d^8$ configuration (i.e., $d^{8-n} \approx d^8\underline{L}^n$) -- was evoked early on~\cite{Mizokawa2000}. Nowadays, the MIT is  usually viewed as a charge ordering of the type $2 (d^8\underline{L}^1) \rightarrow (d^8)+(d^8 \underline{L}^2)$~\cite{Park2012,Johnston2014,Bisogni2016}. In this scenario, $B_{OC}$ appears important to stabilise the charge ordering and open the gap. As suggested in Ref. \cite{De_la_cruz2002} and confirmed from statistical correlation techniques~\cite{Balachandran2013}, $R_{xy}$ and $M_z$ are also expected to play an active role.  However, a complete picture linking electronic, structural and magnetic properties is yet to emerge. 

Unlike recent theoretical studies which were focusing specifically on the electronic properties~\cite{Park2012,Johnston2014,Ruppen2015,Varignon2016},  
we investigate here self-consistently the electronic and structural degrees of freedom of $R$NiO$_3$ compounds from density functional theory calculations 
(DFT, see Methods). Specific care was given to the validation of our approach, which turns out to provide an unprecedented agreement with experimental 
data. Focusing on YNiO$_3$, we show (see Supplementary S1) that not only the atomic structure but also the AFM-E' ground state, the estimated $T_N$ and the 
electronic bandgap of the insulating phase are very accurately reproduced, making therefore our approach a method of choice to shed light on the interlink between 
these different features. 

We start from the reference $Pm\bar{3}m$ cubic structure. Inspection of the phonon dispersion curves (see Supplementary S2) reveals dominant structural 
instabilities at R ($\omega_R =$ 310i cm$^{-1}$)  and M ($\omega_M$ = 278i cm$^{-1}$) points of the Brillouin zone (BZ), which are associated respectively to 
the $R_{xy}$ and $M_{z}$ distortions responsible for the $Pbnm$ phase. These imaginary frequencies $\omega_i$ are linked to a negative energy 
curvature $\alpha_i$ at the origin ($\alpha_i \propto \omega_i^2<0$) and 
to a typical double-well (DW) shape of the energy when freezing $R_{xy}$ and $M_{z}$ distortions of increasing amplitude within the cubic structure, as 
illustrated in Fig. 1. These wells are nearly independent of the spin order but strongly evolve with the $R$ cation : they become shallower when $R$ size 
increases, consistently with a reduction of the related distortion amplitudes in the $Pbnm$ phase. 

In contrast, the $B_{OC}$ motion, corresponding to another phonon at R, is stable and extremely stiff (in fact the stiffest mode with $\omega_B=$ 700 cm$^{-1}$), 
in line with the single well (SW) shape illustrated in Fig. 1. 
Decreasing $R$-cation size tends to stabilise slightly further $B_{OC}$, in apparent contradiction with the observed increase of $T_{MI}$. As illustrated for 
YNiO$_3$,  switching from ferromagnetic (FM) to AFM-E'  spin order reduces slightly the curvature but does not reverse it; instead it shifts the SW to lower 
energy~\cite{Lee2011},  yielding  a finite $Q_B$ at the minimum. Although $B_{OC}$ tends to make the system insulating, this amplitude (25\% of 
ground-state's value)  is however not large enough to open a gap (more than 75\% would be required). This shows that $B_{OC}$ and the magnetic order 
only cannot explain the MIT by themselves.  

\begin{figure*}
\centering
\resizebox{12cm}{!}{\includegraphics{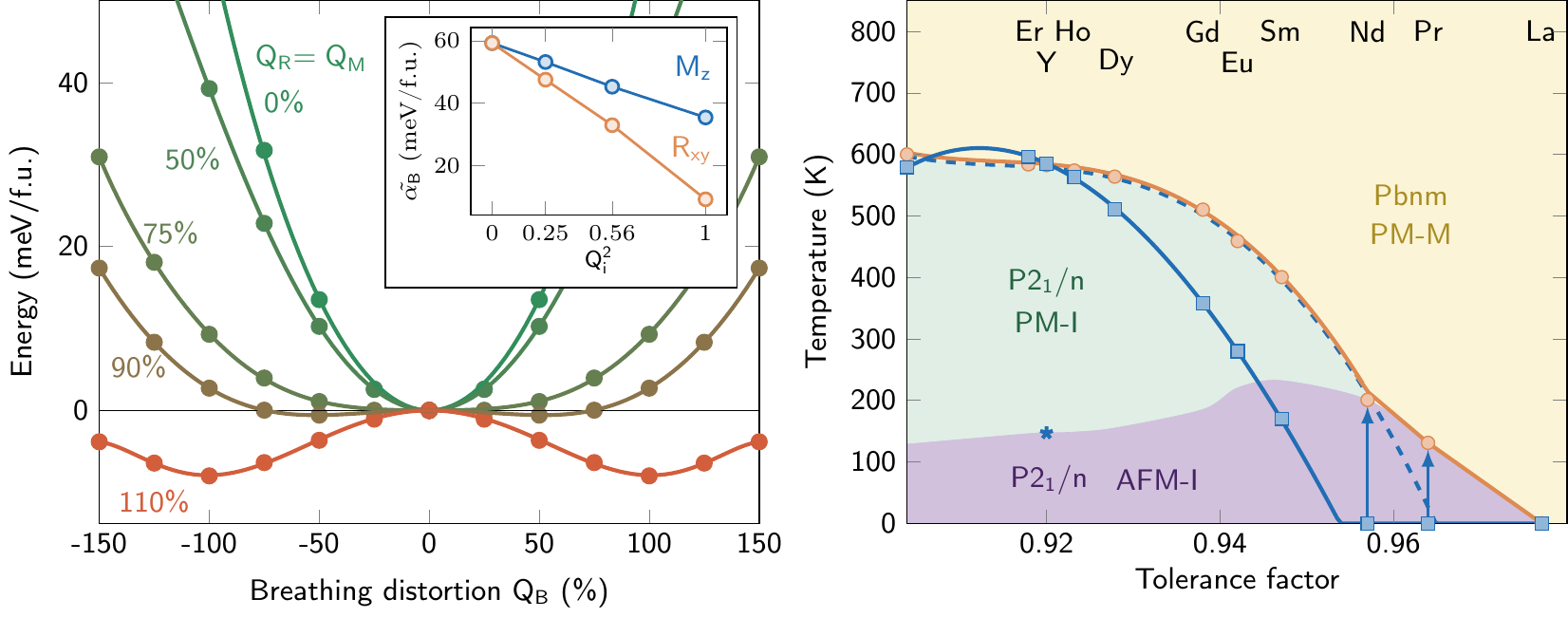}}
\caption{ {\bf Triggered mechanism and nickelate phase diagram} ({\bf a}) Evolution of the energy $E$ in terms of the amplitude of the breathing distortion ($Q_{B}$) for fixed amplitudes of oxygen rotations ($Q_R = Q_M$ from 0\% to 110\%) in the FM cubic phase of YNiO$_3$ (same volume as the ground state). It highlights the softening of the energy well associated to $B_{OC}$, triggered by the oxygen rotations $R_{xy}$ and $M_z$. Inset : Linear evolution of the energy curvature at the origin, $\tilde{\alpha}_B$ along $Q_B$, in terms of the square of the amplitude of the individual distortions $Q_R$ (orange) and $Q_M$ (blue). ({\bf b}) Phase diagram of $R$NiO$_3$ compounds in terms of their tolerance factor $t$ and the temperature $T$. It includes 3 phases : a metallic $Pbnm$ paramagnetic phase (PM-M, yellow area), an insulating $P2_1/n$ paramagnetic phase (PM-I, green area) and an insulating $P2_1/n$ E'-type AFM phase (AFM-I, magenta area). The yellow line and dots show the experimental evolution of $T_{MI}$ with the tolerance factor $t$. The blue line and squares is the prediction of the simple Landau model fitted on the first-principles data (FM order). The dashed blue line is the fit of the Landau expression of $T_{MI}(t)$ on experimental data. The blue star is the magnetic phase transition predicted for YNiO$_3$ from first-principles. The blue arrows indicate the correction to be applied on $T_{MI}$ for large cations  when properly incorporating the change of magnetic order. }
  \label{fig-1}
\end{figure*}

Our central result is presented in Fig. 2 where we report the evolution of the $B_{OC}$ energy well of YNiO$_3$ at various fixed amplitudes of oxygen rotation 
motions. It highlights that, although initially stable (SW), $B_{OC}$ is progressively destabilized (DW) by the appearance of  $R_{xy}$ and $M_z$. As illustrated 
in the inset, $\alpha_B$ is renormalized into $\tilde{\alpha}_B$  which evolves linearly with $Q_R^2$ and $Q_M^2$.  The slope associated to $Q_R$ is twice as 
large as that related to $Q_M$, attesting that each of the three individual rotations similarly affects  $B_{OC}$.  This behavior arises from the presence in the energy 
expansion of cooperative ($\lambda <0$) bi-quadratic coupling terms between $B_{OC}$ and oxygen rotations ($E \approx \lambda_{Bi} Q_{B}^2 Q_i^2$, $i = R, M$)  
which, being the lowest-order couplings allowed by symmetry, should give rise to the appearance of $B_{OC}$ through a "triggered" phase transition according to 
Holakovsky~\cite{Holakovsky1973}. The same behavior is observed independently of the magnetic order (see Supplementary S3). From now we focus on representative FM results while coming back to the role of magnetism later. 
   
To further assess the relevance of such a triggered mechanism in nickelates, we built a Landau model including $R_{xy}$, $M_z$ and $B_{OC}$ degrees of 
freedom~\cite{Balachandran2013}, restricting to lowest coupling terms and assuming temperature dependence of the oxygen rotations only~:
\begin{eqnarray}
  E &=& \gamma_R (T - T_{0R})  Q_{R}^2 + \beta_R Q_{R}^4 + \lambda_{BR} Q_{B}^2 Q_{R}^2  \nonumber \\
  &+&\gamma_M (T - T_{0M})  Q_{M}^2 + \beta_M Q_{M}^4 + \lambda_{BM} Q_{B}^2 Q_{M}^2  \nonumber \\
  &+& \alpha_B  Q_{B}^2 + \beta_B Q_{B}^4  + \lambda_{MR} Q_{M}^2 Q_{R}^2.
\end{eqnarray}
Within this model, $R_{xy}$ and $M_z$ appear at $T_{0R}$ and $T_{0M}$.  On cooling, they progressively develop within the $Pbnm$ phase and renormalize 
the energy curvature $\alpha_B$ of $B_{OC}$ as made clear when grouping the $Q_B^2$ terms  in Eq. (1):
\begin{eqnarray}
  \tilde{\alpha}_B  &=& \alpha_B + \lambda_{BM} Q_{M}^2 + \lambda_{BR} Q_{R}^2   
\end{eqnarray}
When reaching a critical amplitude at which $\tilde{\alpha}_B=0$, they trigger the appearance of $B_{OC}$ and produce concurrent structural and 
metal-insulator phase transitions. The phase transition appears to be second order within this model, which is however too simple to be conclusive 
on this point (see Supplementary S3).

All parameters and their evolution with $R$ were directly fitted from first-principles; only Curie temperatures were uniformly scaled to reproduce the
experimental $T_{MI}$ of YNiO$_3$ (see Supplementary S3). The phase diagram of nickelates as predicted within this model is reported in Fig. 2. 
This figure demonstrates that the cooperative coupling of $B_{OC}$ with  $R_{xy}$ and $M_z$ is a key mechanism that, by itself, accounts 
for the experimentally observed evolution of $T_{MI}$ with the tolerance factor. 

Hence, the MIT in nickelates turns out to be a concrete example of triggered phase transition\cite{Flerov1995,Iwata2007}, a kind of transition never identified before 
in simple perovskites. Indeed, although bi-quadratic interactions are generic in this class of compounds, different distortions usually compete and 
exclude each other \cite{Benedek2013}. The cooperative coupling of $B_{OC}$ with oxygen rotations pointed out here is therefore an unusual 
and intriguing feature, whose origin is now traced back in the electronic band structure.

In the cubic phase, as expected from the formal Ni $3d^7$ ($t_{2g}^6e_g^1$) occupancy, the Fermi energy $E_f$ crosses levels of dominant 
Ni $3d$-$e_g$ character (i.e. anti-bonding Ni $3d$- O $2p$ states); such levels form an isolated and rather dispersive set of two $e_g$ bands, shifted above the $t_{2g}$ 
levels by the crystal field (Fig. 3). Forcing into this cubic structure a $B_{OC}$ distortion, associated to a phonon at $\mathbf{q}_R = (1/2, 1/2, 1/2)$,  can open a gap at $\mathbf{q}_c = (1/4, 1/4, 1/4)$ within the $e_g$ bands but well above $E_f$ and without any direct impact on  the metallic character and the occupied states. Nevertheless, 
the oxygen rotation motions substantially affect the $e_g$ bands (Fig. 3),  reducing their bandwidth and yielding a progressive down shift of the $e_g$ levels at $\mathbf{q}_c$. 
As the rotations gradually increase and the bandwidth decreases, $B_{OC}$ more substantially pushes down the electronic states around $E_f$, providing a progressive gain of electronic energy which, in turn, results in the softening of $\omega_B$. The critical rotation amplitude 
at which $B_{OC}$ becomes unstable ($\tilde{\alpha}=0$) is precisely that at which the $e_g$ levels at $\mathbf{q}_c$ cross $E_f$. At these amplitudes, the electronic system itself becomes unstable; the appearance of $B_{OC}$ is favored and opens a gap precisely at $E_f$, making the system insulating.  As such, the MIT can therefore be interpreted as a Peierls instability but one that is not initially present and has been triggered by oxygen rotations.

For compounds with small $R$ cations, $Q_{R}$ and $Q_M$ are large and able to produce the MIT at relatively high temperatures, well above $T_N$. For large cations 
($R =$ Nd, Pr), oxygen rotations are reduced and, from our Landau model (built on FM results), no more sufficient to trigger the MIT (Figure 2b). 
However, as previously mentioned, the AFM-E' spin order is compatible by symmetry with $B_{OC}$ and induces its appearance as an improper order (linear shift of 
$B_{OC}$ SW, Fig. 1d). Hence, although not opening a gap in the cubic phase, the onset of the AFM-E' order in the $Pbnm$ phase of NdNiO$_3$ and PrNiO$_3$ promotes the occurrence of the MIT almost triggered by the rotations. In these compounds, we have therefore $T_{MI} =T_N$; the transition is more abrupt and first-order \cite{Catalan2008}. Such active role of magnetism for large cations is supported by the experimental results and discussion in  Ref. \cite{Vobornik1999}. It is also confirmed by our first-principles calculations on NdNiO$_3$, showing that the predicted $T_{MI}$ is rescaled when including the change of magnetic order: while the system prefers to stay in the $Pbnm$ metallic phase when imposing a FM order, it switches to the $P2_1/n$ phase when adopting a AFM-E' order. The cooperative effect of the magnetic order remains true for small cations but without any impact on $T_{MI}$ ($> T_N$).  

In conclusion, the concurrent electronic and structural transitions at $T_{MI}$ in $R$NiO$_3$ compounds take the form of a Peierls instability which, primarily, is
structurally triggered by the oxygen rotation motions $R_{xy}$ and $M_z$ and, eventually, is further assisted by the appearance of the E'-type AFM magnetic ordering. 
Our Landau model, and its possible extension to incorporate explicitly strain degrees of freedom neglected here for simplicity,  provides a simple and useful 
quantitative tool to estimate and interpret how $T_{MI}$ can be tuned toward the monitoring of  oxygen rotation motions $R_{xy}$ and $M_z$ when making solid-solutions \cite{Medarde1997}, applying pressure \cite{Catalan2008} or playing with the epitaxial strain and the orientation of the substrate in thin films \cite{Catalano2015}.  Our findings are relevant to other families of perovskites like A$^{2+}$Fe$^{4+}$O$_3$ compounds~\cite{Akao2003}. For instance, they  can explain why CaFeO$_3$, which exhibits oxygen rotations, undergoes a MIT while SrFeO$_3$ and BaFeO$_3$, which stay cubic, remain metallic. In addition, the same physics is also inherent to manganites like LaMnO$_3$, suggesting a close competition between charge and orbital orderings in this family compounds. However, the situation is slightly different in bismutate like BaBiO$_3$, in which $B_{OC}$ is intrinsically unstable in the cubic phase \cite{Lichtenstein1991}.

\begin{figure}
\resizebox{7.5cm}{!}{\includegraphics{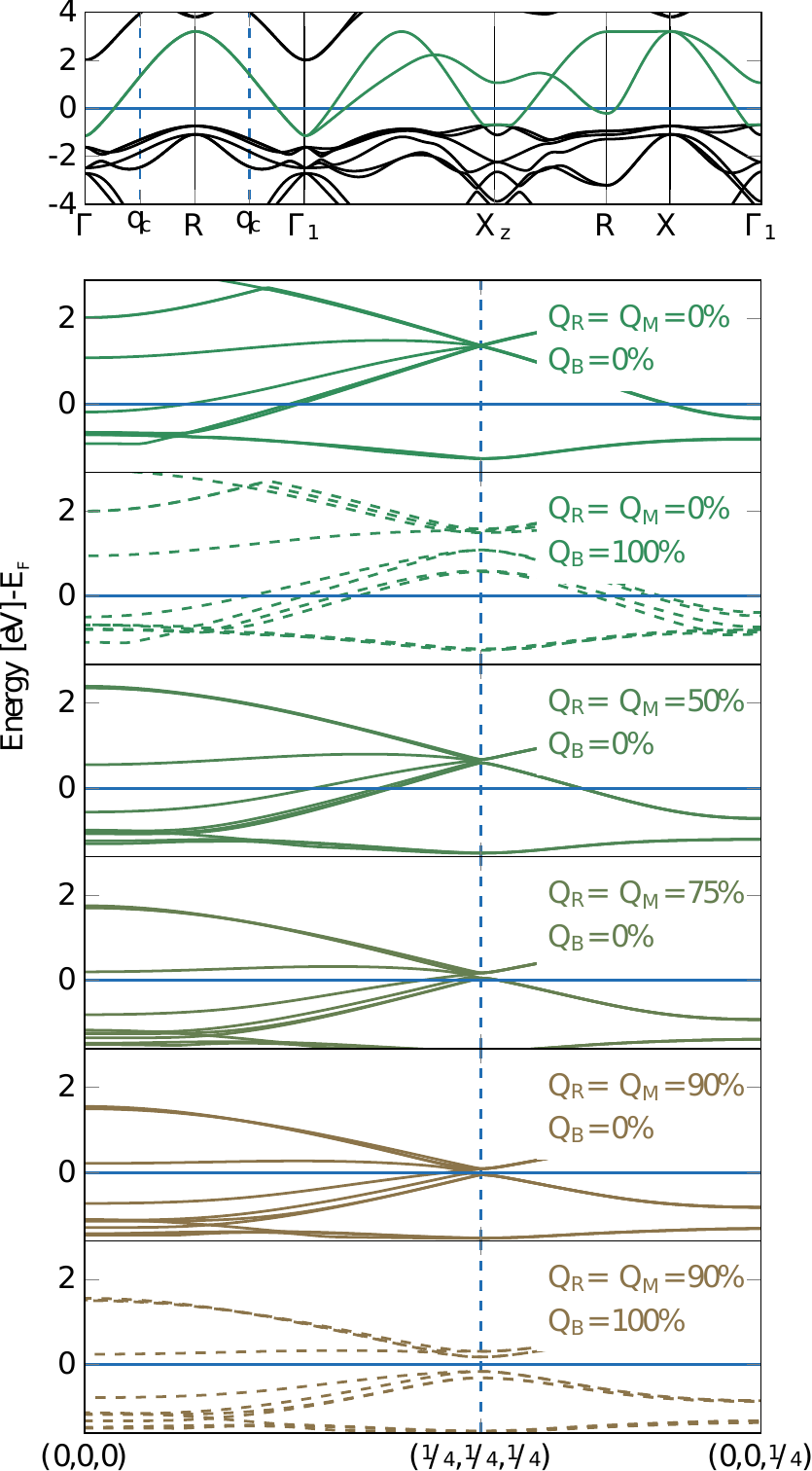}}
  \caption{{\bf Electronic properties} a. Electronic dispersion curves of YNiO$_3$ along different high symmetry line of the Brillouin zone of the $Pm\bar{3}m$ phase (FM case, majority spins):
  $\Gamma = (0,0,0)$, $X = (1/2,0,0)$, $M = (1/2,1/2,0)$ and $R = (1/2,1/2,1/2)$. The Ni 3d $e_g$ bands are highlighted in green. The Fermi energy corresponds to the horizontal blue line. 
  The point $q_c = (1/4, 1/4, 1/4)$ is located by vertical dashed blue lines. (b) Evolution of the electronic dispersion curves around the Fermi energy $E_f$ (FM order, majority spins) when 
  freezing into the $Pm\bar{3}m$ phase increasing amplitudes of oxygen rotations  ($Q_R = Q_M$ from 0\% to 90\%, lines) and eventually adding the breathing distortion ($Q_B = 100$\%, 
  dashed lines). The graph connects high-symmetry points (coordinates in pseudocubic notations) in the Brillouin zone of the $Pbnm$ or $P2_1/n$ 20-atom cell, in which bands have been 
  folded respect to (a). }
  \label{fig:3}
\end{figure}

\section{\label{sec:level10}Methods}

First-principles calculations were performed in the framework of Density Functional Theory (DFT) \cite{Hohenberg1964,Kohn1965}
using a Projected Augmented Wave (PAW) approach \cite{Blochl1994} as implemented within the ABINIT package \cite{Gonze2009,Gonze2002,Gonze2005,Torrent2008}.
The calculations relied on the Generalized Gradient Approximation using the PBEsol \cite{Perdew2008}  exchange-correlation 
functional. We worked within a collinear spin approximation. We included a Hubbard correction U = 1.5 eV \cite{Liechtenstein1995} on the $3d$ orbitals of 
Ni atoms. A special care as been devoted to the determination of the appropriate U parameter (see Supplementary, Section S1).

We made use of JTH atomic potentials \cite{Jollet2014}.  For the wave-functions, we used an energy cutoff of 24 Ha  (38 Ha for the second 
grid in the PAW spheres), which guarantees a convergence better than 1 meV on the total energy. The Brillouin-zone was sampled with $k$-point meshes equivalent 
to a $12 \times 12 \times 12$ grid in the 5-atoms unit cell. During structural relaxations, thresholds of $10^{-5}$ Ha/bohr on the maximum force and of $10^{-7}$ 
Ha/bohr$^3$ on the maximum stress have been considered.

The Goldschmidt tolerance factor, \cite{Goldschmidt1926}, $t = d_{R-O} / \sqrt{2} d_{Ni-O}$, of $R$NiO$_3$ compounds were determined using Nicole Benedek's tool \cite{Benedek-tol} relying 
on a bond valence model \cite{Lufaso2001} to calculate $d_{R-O}$ and $d_{Ni-O}$, respectively the ideal $R$--O and Ni--O bond lengths in the cubic perovskite structure. 
 
Symmetry-adapted mode analysis have been performed with AMPLIMODE \cite{Orobengoa2009,Perez-Mato2010}. The modes are normalized to their amplitude in the $P2_1/n$ AFM-E' ground state. This normalization is such that in cubic phase (volume of the ground state) $Q_R = 100\%$ corresponds to rotation angles  $\phi_x = \phi_y =$ 11.33$^o$ (Ni-O-Ni angle of 157.33$^o$) , $Q_R = 100\%$ corresponds to a rotation angle $\phi_z =$ 12.12$^o$ (Ni-O-Ni angle of 155.75$^o$)
and $Q_B = 100\% $ corresponds to oxygen displacements $d_O = 0.0358 \AA$.

The Landau model parameters have been fitted for  YNiO$_3$, GdNiO$_3$ and SmNiO$_3$ on first-principles data using in each case a FM cubic phase (volume of the ground state) and interpolated for the other compounds.  $T_{MI}$ was determined analytically (see Supplementary, Section S3).

\section{\label{sec:level10}Supplementary Information}
Supplementary Information is available in the online version of the paper.

\section{\label{sec:level10}Acknowledgements}
The authors thank Y.~Zhang, E.~Bousquet, F.~Ricci, M.~Verstraete, J.-Y. Raty, S. Catalano and J. Varignon for fruitful discussions.
This work was supported by FRS-FNRS project HiT4FiT and ARC project AIMED. J.I. acknowledges support form FNR 
Luxembourg Grant P12/4853155 "COFERMAT". Calculations have
been performed  on the C\'eci facilities funded by F.R.S-FNRS (Grant No 2.5020.1) and Tier-1 supercomputer 
of the F\'ed\'eration Wallonie-Bruxelles funded by the Walloon Region (Grant No 1117545). They also took 
advantage of HPC resources from the French Research and Technology Computing Center (CCRT) and from
the PRACE project Megapasta.

\bibliography{biblio}

\begin{thebibliography}{10}
\expandafter\ifx\csname url\endcsname\relax
  \def\url#1{\texttt{#1}}\fi
\expandafter\ifx\csname urlprefix\endcsname\relax\def\urlprefix{URL }\fi
\providecommand{\bibinfo}[2]{#2}
\providecommand{\eprint}[2][]{\url{#2}}

\bibitem{Demazeau1971}
\bibinfo{author}{Demazeau, G.}, \bibinfo{author}{Marbeuf, A.},
  \bibinfo{author}{Pouchard, M.} \& \bibinfo{author}{Hagenmuller, P.}
\newblock \bibinfo{title}{Sur une s\'erie de compos\'es oxyg\`enes du nickel
  trivalent deriv\'es de la perovskite}.
\newblock \emph{\bibinfo{journal}{J. Solid State Chem.}}
  \textbf{\bibinfo{volume}{3}}, \bibinfo{pages}{582} (\bibinfo{year}{1971}).

\bibitem{Shi2013}
\bibinfo{author}{Shi, J.}, \bibinfo{author}{Ha, S.~D.}, \bibinfo{author}{Zhou,
  Y.}, \bibinfo{author}{Schoofs, F.} \& \bibinfo{author}{Ramanathan, S.}
\newblock \bibinfo{title}{A correlated nickelate synaptic transistor}.
\newblock \emph{\bibinfo{journal}{Nature Communications}}
  \textbf{\bibinfo{volume}{4}}, \bibinfo{pages}{2676} (\bibinfo{year}{2013}).

\bibitem{Zhou2016}
\bibinfo{author}{Zhou, Y.} \emph{et~al.}
\newblock \bibinfo{title}{Strongly correlated perovskite fuel cells}.
\newblock \emph{\bibinfo{journal}{Nature}} \textbf{\bibinfo{volume}{534}},
  \bibinfo{pages}{231--234} (\bibinfo{year}{2016}).

\bibitem{Middey2016}
\bibinfo{author}{Middey, S.} \emph{et~al.}
\newblock \bibinfo{title}{Physics of {Ultrathin} {Films} and {Heterostructures}
  of {Rare}-{Earth} {Nickelates}}.
\newblock \emph{\bibinfo{journal}{Annual Review of Materials Research}}
  \textbf{\bibinfo{volume}{46}}, \bibinfo{pages}{305--334}
  (\bibinfo{year}{2016}).

\bibitem{Forst2015}
\bibinfo{author}{Först, M.} \emph{et~al.}
\newblock \bibinfo{title}{Spatially resolved ultrafast magnetic dynamics
  initiated at a complex oxide heterointerface}.
\newblock \emph{\bibinfo{journal}{Nature Materials}}
  \textbf{\bibinfo{volume}{14}}, \bibinfo{pages}{883--888}
  (\bibinfo{year}{2015}).

\bibitem{Kim2016}
\bibinfo{author}{Kim, T.~H.} \emph{et~al.}
\newblock \bibinfo{title}{Polar metals by geometric design}.
\newblock \emph{\bibinfo{journal}{Nature}} \textbf{\bibinfo{volume}{533}},
  \bibinfo{pages}{68--72} (\bibinfo{year}{2016}).

\bibitem{Grisolia2016}
\bibinfo{author}{Grisolia, M.~N.} \emph{et~al.}
\newblock \bibinfo{title}{Hybridization-controlled charge transfer and induced
  magnetism at correlated oxide interfaces}.
\newblock \emph{\bibinfo{journal}{Nature Physics}}
  \textbf{\bibinfo{volume}{12}}, \bibinfo{pages}{484--492}
  (\bibinfo{year}{2016}).

\bibitem{Catalan2008}
\bibinfo{author}{Catalan, G.}
\newblock \bibinfo{title}{Progress in perovskite nickelate research}.
\newblock \emph{\bibinfo{journal}{Phase Transitions}}
  \textbf{\bibinfo{volume}{81}}, \bibinfo{pages}{729--749}
  (\bibinfo{year}{2008}).

\bibitem{Benedek2013}
\bibinfo{author}{Benedek, N.~A.} \& \bibinfo{author}{Fennie, C.~J.}
\newblock \bibinfo{title}{Why are there so few perovskite ferroelectrics?}
\newblock \emph{\bibinfo{journal}{The Journal of Physical Chemistry C}}
  \textbf{\bibinfo{volume}{117}}, \bibinfo{pages}{13339--13349}
  (\bibinfo{year}{2013}).
\newblock \eprint{http://dx.doi.org/10.1021/jp402046t}.

\bibitem{Medarde1997}
\bibinfo{author}{Medarde, M.~L.}
\newblock \bibinfo{title}{Structural, magnetic and electronic properties of
  {RNiO$_3$} perovskites ({R} = rare earth)}.
\newblock \emph{\bibinfo{journal}{Journal of Physics: Condensed Matter}}
  \textbf{\bibinfo{volume}{9}}, \bibinfo{pages}{1679} (\bibinfo{year}{1997}).

\bibitem{Alonso1999}
\bibinfo{author}{Alonso, J.~A.} \emph{et~al.}
\newblock \bibinfo{title}{Charge {Disproportionation} in {$RNiO_3$}
  {Perovskites}: {Simultaneous} {Metal}-{Insulator} and {Structural}
  {Transition} in {$YNiO3$}}.
\newblock \emph{\bibinfo{journal}{Physical Review Letters}}
  \textbf{\bibinfo{volume}{82}}, \bibinfo{pages}{3871--3874}
  (\bibinfo{year}{1999}).

\bibitem{Garcia-munoz1994}
\bibinfo{author}{García-Muñoz, J.~L.}, \bibinfo{author}{Rodríguez-Carvajal,
  J.} \& \bibinfo{author}{Lacorre, P.}
\newblock \bibinfo{title}{Neutron-diffraction study of the magnetic ordering in
  the insulating regime of the perovskites {RNiO$_3$} {R} ={Pr} and {Nd})}.
\newblock \emph{\bibinfo{journal}{Physical Review B}}
  \textbf{\bibinfo{volume}{50}}, \bibinfo{pages}{978--992}
  (\bibinfo{year}{1994}).

\bibitem{Lee2011}
\bibinfo{author}{Lee, S.}, \bibinfo{author}{Chen, R.} \&
  \bibinfo{author}{Balents, L.}
\newblock \bibinfo{title}{Landau {Theory} of {Charge} and {Spin} {Ordering} in
  the {Nickelates}}.
\newblock \emph{\bibinfo{journal}{Physical Review Letters}}
  \textbf{\bibinfo{volume}{106}}, \bibinfo{pages}{016405}
  (\bibinfo{year}{2011}).

\bibitem{Torrance1992}
\bibinfo{author}{Torrance, J.~B.}, \bibinfo{author}{Lacorre, P.},
  \bibinfo{author}{Nazzal, A.~I.}, \bibinfo{author}{Ansaldo, E.~J.} \&
  \bibinfo{author}{Niedermayer, C.}
\newblock \bibinfo{title}{Systematic study of insulator-metal transitions in
  perovskites {RNiO$_3$} {R} ={Pr},{Nd},{Sm},{Eu}) due to closing of
  charge-transfer gap}.
\newblock \emph{\bibinfo{journal}{Physical Review B}}
  \textbf{\bibinfo{volume}{45}}, \bibinfo{pages}{8209--8212}
  (\bibinfo{year}{1992}).

\bibitem{Mizokawa2000}
\bibinfo{author}{Mizokawa, T.}, \bibinfo{author}{Khomskii, D.~I.} \&
  \bibinfo{author}{Sawatzky, G.~A.}
\newblock \bibinfo{title}{Spin and charge ordering in self-doped {Mott}
  insulators}.
\newblock \emph{\bibinfo{journal}{Physical Review B}}
  \textbf{\bibinfo{volume}{61}}, \bibinfo{pages}{11263--11266}
  (\bibinfo{year}{2000}).

\bibitem{Raebiger2008}
\bibinfo{author}{Raebiger, H.}, \bibinfo{author}{Lany, S.} \&
  \bibinfo{author}{Zunger, A.}
\newblock \bibinfo{title}{Charge self-regulation upon changing the oxidation
  state of transition metals in insulators}.
\newblock \emph{\bibinfo{journal}{Nature}} \textbf{\bibinfo{volume}{453}},
  \bibinfo{pages}{763--766} (\bibinfo{year}{2008}).

\bibitem{Park2012}
\bibinfo{author}{Park, H.}, \bibinfo{author}{Millis, A.~J.} \&
  \bibinfo{author}{Marianetti, C.~A.}
\newblock \bibinfo{title}{Site-{Selective} {Mott} {Transition} in
  {Rare}-{Earth}-{Element} {Nickelates}}.
\newblock \emph{\bibinfo{journal}{Physical Review Letters}}
  \textbf{\bibinfo{volume}{109}}, \bibinfo{pages}{156402}
  (\bibinfo{year}{2012}).

\bibitem{Ruppen2015}
\bibinfo{author}{Ruppen, J.} \emph{et~al.}
\newblock \bibinfo{title}{Optical spectroscopy and the nature of the insulating
  state of rare-earth nickelates}.
\newblock \emph{\bibinfo{journal}{Physical Review B}}
  \textbf{\bibinfo{volume}{92}}, \bibinfo{pages}{155145}
  (\bibinfo{year}{2015}).

\bibitem{Varignon2016}
\bibinfo{author}{Varignon, J.}, \bibinfo{author}{Grisolia, M.~N.},
  \bibinfo{author}{Íñiguez, J.}, \bibinfo{author}{Barthélémy, A.} \&
  \bibinfo{author}{Bibes, M.}
\newblock \bibinfo{title}{Reconciling the ionic and covalent pictures in
  rare-earth nickelates}.
\newblock \emph{\bibinfo{journal}{arXiv:1603.05480 [cond-mat]}}
  (\bibinfo{year}{2016}).
\newblock \bibinfo{note}{ArXiv: 1603.05480}.

\bibitem{Mazin2007}
\bibinfo{author}{Mazin, I.~I.} \emph{et~al.}
\newblock \bibinfo{title}{Charge {Ordering} as {Alternative} to {Jahn}-{Teller}
  {Distortion}}.
\newblock \emph{\bibinfo{journal}{Physical Review Letters}}
  \textbf{\bibinfo{volume}{98}}, \bibinfo{pages}{176406}
  (\bibinfo{year}{2007}).

\bibitem{Johnston2014}
\bibinfo{author}{Johnston, S.}, \bibinfo{author}{Mukherjee, A.},
  \bibinfo{author}{Elfimov, I.}, \bibinfo{author}{Berciu, M.} \&
  \bibinfo{author}{Sawatzky, G.~A.}
\newblock \bibinfo{title}{Charge {Disproportionation} without {Charge}
  {Transfer} in the {Rare}-{Earth}-{Element} {Nickelates} as a {Possible}
  {Mechanism} for the {Metal}-{Insulator} {Transition}}.
\newblock \emph{\bibinfo{journal}{Physical Review Letters}}
  \textbf{\bibinfo{volume}{112}}, \bibinfo{pages}{106404}
  (\bibinfo{year}{2014}).

\bibitem{Bisogni2016}
\bibinfo{author}{Bisogni, V.} \emph{et~al.}
\newblock \bibinfo{title}{Ground state oxygen holes and the metal-insulator
  transition in the negative charge transfer rare-earth nickelates}.
\newblock \emph{\bibinfo{journal}{Nature Communications}}
  \textbf{\bibinfo{volume}{7}}, \bibinfo{pages}{13017} (\bibinfo{year}{2016}).
\newblock \bibinfo{note}{ArXiv: 1607.06758}.

\bibitem{De_la_cruz2002}
\bibinfo{author}{de~la Cruz, F.~P.} \emph{et~al.}
\newblock \bibinfo{title}{Possible common ground for the metal-insulator phase
  transition in the rare-earth nickelates {RNiO$_3$} ({R}={Eu}, {Ho}, {Y})}.
\newblock \emph{\bibinfo{journal}{Physical Review B}}
  \textbf{\bibinfo{volume}{66}}, \bibinfo{pages}{153104}
  (\bibinfo{year}{2002}).

\bibitem{Balachandran2013}
\bibinfo{author}{Balachandran, P.~V.} \& \bibinfo{author}{Rondinelli, J.~M.}
\newblock \bibinfo{title}{Interplay of octahedral rotations and breathing
  distortions in charge-ordering perovskite oxides}.
\newblock \emph{\bibinfo{journal}{Physical Review B}}
  \textbf{\bibinfo{volume}{88}}, \bibinfo{pages}{054101}
  (\bibinfo{year}{2013}).

\bibitem{Holakovsky1973}
\bibinfo{author}{Holakovsky, J.}
\newblock \bibinfo{title}{A new type of ferroelectric phase transition}.
\newblock \emph{\bibinfo{journal}{Phys. Stat. Sol. (b)}}
  \textbf{\bibinfo{volume}{56}}, \bibinfo{pages}{615--619}
  (\bibinfo{year}{1973}).

\bibitem{Flerov1995}
\bibinfo{author}{Flerov, I.~N.}, \bibinfo{author}{Gorev, M.~V.},
  \bibinfo{author}{Voronov, V.~N.}, \bibinfo{author}{Tressaud, A.} \&
  \bibinfo{author}{Grannec, J.}
\newblock \bibinfo{title}{Triggered phase transitions in
  {Rb$_2$KB$^{3+}$F$_6$(B$^{3+}$}: Er, ho) elpasolites}.
\newblock \emph{\bibinfo{journal}{J. Solid State Chem.}}
  \textbf{\bibinfo{volume}{3}}, \bibinfo{pages}{582} (\bibinfo{year}{1971}).

\bibitem{Iwata2007}
\bibinfo{author}{Iwata, M.}, \bibinfo{author}{Zhao, C.~H.},
  \bibinfo{author}{Aoyagi, R.}, \bibinfo{author}{Maeda, M.} \&
  \bibinfo{author}{Ishibashi, Y.}
\newblock \bibinfo{title}{Splitting of triggered phase transition in
  {Bi$_{4-x}$La$_x$Ti$_3$O$_{12}$} mixed crystals}.
\newblock \emph{\bibinfo{journal}{Japonese Journal of Applied Physics}}
  \textbf{\bibinfo{volume}{46}}, \bibinfo{pages}{5894} (\bibinfo{year}{2007}).

\bibitem{Vobornik1999}
\bibinfo{author}{Vobornik, I.} \emph{et~al.}
\newblock \bibinfo{title}{Electronic-structure evolution through the
  metal-insulator transition in {$RNiO_3$}}.
\newblock \emph{\bibinfo{journal}{Phys. Rev. B}} \textbf{\bibinfo{volume}{60}},
  \bibinfo{pages}{R8426--R8429} (\bibinfo{year}{1999}).

\bibitem{Catalano2015}
\bibinfo{author}{Catalano, S.} \emph{et~al.}
\newblock \bibinfo{title}{Tailoring the electronic transitions of {NdNiO}3
  films through (111)pc oriented interfaces}.
\newblock \emph{\bibinfo{journal}{APL Materials}} \textbf{\bibinfo{volume}{3}},
  \bibinfo{pages}{062506} (\bibinfo{year}{2015}).

\bibitem{Akao2003}
\bibinfo{author}{Akao, T.} \emph{et~al.}
\newblock \bibinfo{title}{Charge-ordered state in single-crystalline
  {CaFeO$_{3}$} thin film studied by x-ray anomalous diffraction}.
\newblock \emph{\bibinfo{journal}{Phys. Rev. Lett.}}
  \textbf{\bibinfo{volume}{91}}, \bibinfo{pages}{156405}
  (\bibinfo{year}{2003}).

\bibitem{Lichtenstein1991}
\bibinfo{title}{Structural phase diagram and electron-phonon interaction in
  {Ba$_{1-x}$K$_{x}$BiO$_{3}$}, author = {Liechtenstein, A. I. and Mazin, I. I.
  and Rodriguez, C. O. and Jepsen, O. and Andersen, O. K. and Methfessel, M.},
  journal = {Phys. Rev. B}, volume = {44}, issue = {10}, pages = {5388--5391},
  numpages = {0}, year = {1991}, month = {Sep}, publisher = {American Physical
  Society}, doi = {10.1103/PhysRevB.44.5388}, url =
  {https://link.aps.org/doi/10.1103/PhysRevB.44.5388}} .

\bibitem{Hohenberg1964}
\bibinfo{author}{Hohenberg, P.} \& \bibinfo{author}{Kohn, W.}
\newblock \bibinfo{title}{Inhomogeneous electron gas}.
\newblock \emph{\bibinfo{journal}{Phys. Rev.}} \textbf{\bibinfo{volume}{136}},
  \bibinfo{pages}{B864--B871} (\bibinfo{year}{1964}).

\bibitem{Kohn1965}
\bibinfo{author}{Kohn, W.} \& \bibinfo{author}{Sham, L.~J.}
\newblock \bibinfo{title}{Self-consistent equations including exchange and
  correlation effects}.
\newblock \emph{\bibinfo{journal}{Phys. Rev.}} \textbf{\bibinfo{volume}{140}},
  \bibinfo{pages}{A1133--A1138} (\bibinfo{year}{1965}).

\bibitem{Blochl1994}
\bibinfo{author}{Bl\"ochl, P.~E.}
\newblock \bibinfo{title}{Projector augmented-wave method}.
\newblock \emph{\bibinfo{journal}{Phys. Rev. B}} \textbf{\bibinfo{volume}{50}},
  \bibinfo{pages}{17953--17979} (\bibinfo{year}{1994}).

\bibitem{Gonze2009}
\bibinfo{author}{Gonze, X.} \emph{et~al.}
\newblock \bibinfo{title}{{ABINIT}: {First}-principles approach to material and
  nanosystem properties}.
\newblock \emph{\bibinfo{journal}{Computer Physics Communications}}
  \textbf{\bibinfo{volume}{180}}, \bibinfo{pages}{2582--2615}
  (\bibinfo{year}{2009}).

\bibitem{Gonze2002}
\bibinfo{author}{Gonze, X.} \emph{et~al.}
\newblock \bibinfo{title}{First-principles computation of material properties:
  the {ABINIT} software project}.
\newblock \emph{\bibinfo{journal}{Computational Materials Science}}
  \textbf{\bibinfo{volume}{25}}, \bibinfo{pages}{478 -- 492}
  (\bibinfo{year}{2002}).

\bibitem{Gonze2005}
\bibinfo{author}{Gonze, X.} \emph{et~al.}
\newblock \bibinfo{title}{A brief introduction to the abinit software package}
  .

\bibitem{Torrent2008}
\bibinfo{author}{Torrent, M.}, \bibinfo{author}{Jollet, F.},
  \bibinfo{author}{Bottin, F.}, \bibinfo{author}{Zérah, G.} \&
  \bibinfo{author}{Gonze, X.}
\newblock \bibinfo{title}{Implementation of the projector augmented-wave method
  in the {ABINIT} code: {Application} to the study of iron under pressure}.
\newblock \emph{\bibinfo{journal}{Computational Materials Science}}
  \textbf{\bibinfo{volume}{42}}, \bibinfo{pages}{337--351}
  (\bibinfo{year}{2008}).

\bibitem{Perdew2008}
\bibinfo{author}{Perdew, J.~P.} \emph{et~al.}
\newblock \bibinfo{title}{Restoring the density-gradient expansion for exchange
  in solids and surfaces}.
\newblock \emph{\bibinfo{journal}{Phys. Rev. Lett.}}
  \textbf{\bibinfo{volume}{100}}, \bibinfo{pages}{136406}
  (\bibinfo{year}{2008}).

\bibitem{Liechtenstein1995}
\bibinfo{author}{Liechtenstein, A.~I.}, \bibinfo{author}{Anisimov, V.~I.} \&
  \bibinfo{author}{Zaanen, J.}
\newblock \bibinfo{title}{Density-functional theory and strong interactions:
  Orbital ordering in mott-hubbard insulators}.
\newblock \emph{\bibinfo{journal}{Phys. Rev. B}} \textbf{\bibinfo{volume}{52}},
  \bibinfo{pages}{R5467--R5470} (\bibinfo{year}{1995}).

\bibitem{Jollet2014}
\bibinfo{author}{Jollet, F.}, \bibinfo{author}{Torrent, M.} \&
  \bibinfo{author}{Holzwarth, N.}
\newblock \bibinfo{title}{Generation of {Projector} {Augmented}-{Wave} atomic
  data: {A} 71 element validated table in the {XML} format}.
\newblock \emph{\bibinfo{journal}{Computer Physics Communications}}
  \textbf{\bibinfo{volume}{185}}, \bibinfo{pages}{1246--1254}
  (\bibinfo{year}{2014}).

\bibitem{Goldschmidt1926}
\bibinfo{author}{Goldschmidt, V.~M.}
\newblock \bibinfo{title}{Die gesetze der krystallochemie}.
\newblock \emph{\bibinfo{journal}{Naturwissenschaften}}
  \textbf{\bibinfo{volume}{14}}, \bibinfo{pages}{477--485}
  (\bibinfo{year}{1926}).

\bibitem{Benedek-tol}
\bibinfo{author}{Benedek, N.}
\newblock \bibinfo{title}{Tolerance factor calculator v0.2.1} .

\bibitem{Lufaso2001}
\bibinfo{author}{Lufaso, M.~W.} \& \bibinfo{author}{Woodward, P.~M.}
\newblock \bibinfo{title}{Prediction of the crystal structures of perovskites
  using the software program spuds}.
\newblock \emph{\bibinfo{journal}{Acta Crystallogr. B}}
  \textbf{\bibinfo{volume}{57}}, \bibinfo{pages}{725} (\bibinfo{year}{2001}).

\bibitem{Orobengoa2009}
\bibinfo{author}{Orobengoa, D.}, \bibinfo{author}{Capillas, C.},
  \bibinfo{author}{Aroyo, M.~I.} \& \bibinfo{author}{Perez-Mato, J.~M.}
\newblock \bibinfo{title}{{{\it AMPLIMODES}: symmetry-mode analysis on the
  Bilbao Crystallographic Server}}.
\newblock \emph{\bibinfo{journal}{Journal of Applied Crystallography}}
  \textbf{\bibinfo{volume}{42}}, \bibinfo{pages}{820--833}
  (\bibinfo{year}{2009}).

\bibitem{Perez-Mato2010}
\bibinfo{author}{Perez-Mato, J.~M.}, \bibinfo{author}{Orobengoa, D.} \&
  \bibinfo{author}{Aroyo, M.~I.}
\newblock \bibinfo{title}{{Mode crystallography of distorted structures}}.
\newblock \emph{\bibinfo{journal}{Acta Crystallographica Section A}}
  \textbf{\bibinfo{volume}{66}}, \bibinfo{pages}{558--590}
  (\bibinfo{year}{2010}).

\bibitem{Alonso2001}
\bibinfo{author}{Alonso, J.~A.} \emph{et~al.}
\newblock \bibinfo{title}{High-temperature structural evolution of {R} {NiO} 3
  ( {R} = {H} o , {Y} , {E} r , {Lu} ) perovskites: {Charge} disproportionation
  and electronic localization}.
\newblock \emph{\bibinfo{journal}{Physical Review B}}
  \textbf{\bibinfo{volume}{64}} (\bibinfo{year}{2001}).

\bibitem{Prosandeev2012}
\bibinfo{author}{Prosandeev, S.}, \bibinfo{author}{Bellaiche, L.} \&
  \bibinfo{author}{Íñiguez, J.}
\newblock \bibinfo{title}{Ab initio study of the factors affecting the ground
  state of rare-earth nickelates}.
\newblock \emph{\bibinfo{journal}{Physical Review B}}
  \textbf{\bibinfo{volume}{85}}, \bibinfo{pages}{214431}
  (\bibinfo{year}{2012}).

\bibitem{Alonso2009}
\bibinfo{author}{Muñoz, A.}, \bibinfo{author}{Alonso, J.~A.},
  \bibinfo{author}{Martínez-Lope, M.~J.} \& \bibinfo{author}{Fernández-Díaz,
  M.~T.}
\newblock \bibinfo{title}{On the magnetic structure of {DyNiO}3}.
\newblock \emph{\bibinfo{journal}{Journal of Solid State Chemistry}}
  \textbf{\bibinfo{volume}{182}}, \bibinfo{pages}{1982--1989}
  (\bibinfo{year}{2009}).

\bibitem{Mercy2017}
\bibinfo{author}{Mercy, A.}, \bibinfo{author}{J., B.} \&
  \bibinfo{author}{Ghosez, P.}
\newblock \bibinfo{title}{Magnetic properties of {YNiO$_3$}}.
\newblock \emph{\bibinfo{journal}{unpublished}}  (\bibinfo{year}{2017}).

\bibitem{Arima1993}
\bibinfo{author}{Arima, T.}, \bibinfo{author}{Tokura, Y.} \&
  \bibinfo{author}{Torrance, J.~B.}
\newblock \bibinfo{title}{Variation of optical gaps in perovskite-type 3
  \textit{ d } transition-metal oxides}.
\newblock \emph{\bibinfo{journal}{Physical Review B}}
  \textbf{\bibinfo{volume}{48}}, \bibinfo{pages}{17006--17009}
  (\bibinfo{year}{1993}).

\bibitem{Perez1999}
\bibinfo{author}{Perez-Cacho, J.}, \bibinfo{author}{Blasco, J.},
  \bibinfo{author}{Garcia, J.}, \bibinfo{author}{Castro, M.} \&
  \bibinfo{author}{Stankiewicz, J.}
\newblock \bibinfo{title}{Study of the phase transitions in {SmNiO$_3$}}.
\newblock \emph{\bibinfo{journal}{J. Phys.: Condens. Matter}}
  \textbf{\bibinfo{volume}{11}}, \bibinfo{pages}{405} (\bibinfo{year}{1999}).

\bibitem{Nikulina2004}
\bibinfo{author}{Nikulina, I.}, \bibinfo{author}{Novojilova, M.},
  \bibinfo{author}{Kaulb, A.}, \bibinfo{author}{Maiorovab, A.} \&
  \bibinfo{author}{S.N., M.}
\newblock \bibinfo{title}{Synthesis and transport properties study of
  {Nd$_{1-x}$Sm$_x$NiO$_{3-\delta}$} solid solutions}.
\newblock \emph{\bibinfo{journal}{Materials Research Bulletin}}
  \textbf{\bibinfo{volume}{39}}, \bibinfo{pages}{803} (\bibinfo{year}{2004}).

\bibitem{Iniguez2001}
\bibinfo{author}{\'I\~niguez, J.}, \bibinfo{author}{Ivantchev, S.},
  \bibinfo{author}{Perez-Mato, J.~M.} \& \bibinfo{author}{Garc\'{\i}a, A.}
\newblock \bibinfo{title}{Devonshire-landau free energy of {BaTiO$_{3}$} from
  first principles}.
\newblock \emph{\bibinfo{journal}{Phys. Rev. B}} \textbf{\bibinfo{volume}{63}},
  \bibinfo{pages}{144103} (\bibinfo{year}{2001}).

\bibitem{Zhong1995}
\bibinfo{author}{Zhong, W.}, \bibinfo{author}{Vanderbilt, D.} \&
  \bibinfo{author}{Rabe, K.~M.}
\newblock \bibinfo{title}{First-principles theory of ferroelectric phase
  transitions for perovskites: The case of {BaTiO$_{3}$}}.
\newblock \emph{\bibinfo{journal}{Phys. Rev. B}} \textbf{\bibinfo{volume}{52}},
  \bibinfo{pages}{6301--6312} (\bibinfo{year}{1995}).

\end{thebibliography}

\onecolumngrid
\newpage
\section{-- {\it Supplementary Information} -- \\ Structurally Triggered Metal-Insulator Transition \\ in Rare-Earth Nickelates   }
\section*{\label{sec:levelsupp1}S1. Validation of the DFT+U approach}

In order to assess the validity of our DFT+U approach and determine the appropriate U parameter, we have considered a wide range of possible values 
for U (from 0 to 8 eV) and have compared the computed structural, magnetic and electronic properties to experimental data. 

The results are summarized below for YNiO$_3$ considered as a test case. In line with what was reported independently in Ref. \cite{Varignon2016}, 
it appears that a DFT approach with a moderate U value of 1.5 eV provides for nickelates an unprecedented agreement with experimental data, combining 
accurate description not only of the structural but also of the magnetic and electronic properties. It therefore offers a robust and ideal framework 
for the study of the interplay between these properties. 

\subsection{\label{subsec1:levelsupp1}A. Atomic structure}

In Figure S1, we report the relative deviations respect to experimental data at low temperature \cite{Alonso2001} for the lattice parameters 
and atomic distortions in the E'-type AFM $P2_1/n$ phase of YNiO$_3$  in terms of the amplitude of the U parameter. The atomic distortions 
are those with respect to the $Pm\bar{3}m$ phase and are quantified from a symmetry-adapted mode analysis performed with 
AMPLIMODE \cite{Orobengoa2009,Perez-Mato2010}.  The labels of the modes that are allowed by symmetry in the $Pbnm$ and $P2_1/n$ phases 
and a brief description of the related atomic motions are reported in Table S1. 

\begin{figure}[h!]
 \includegraphics[width=0.8\linewidth]{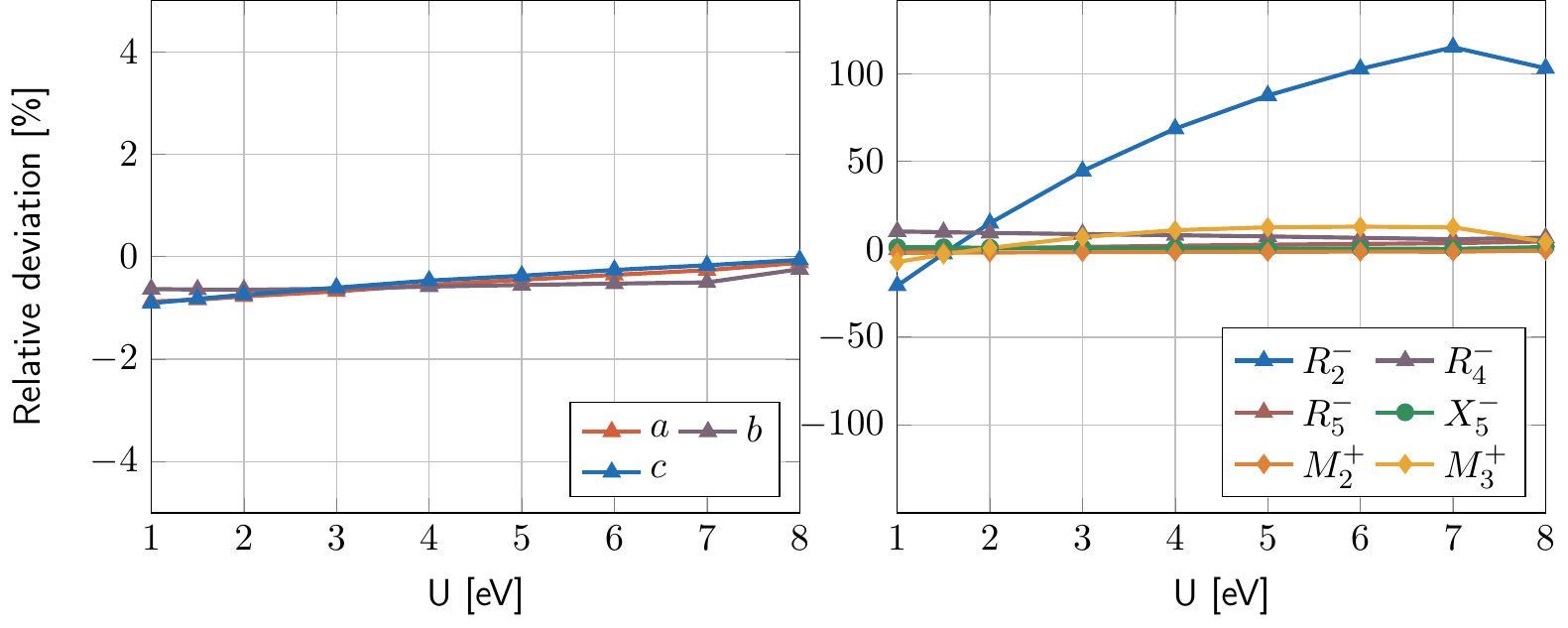}  
 \caption{Relative deviations respect to experimental data at low temperature \cite{Alonso2001} for the lattice parameters (left panel) 
and atomic distortions (right panel) in the E'-type AFM $P2_1/n$ phase of YNiO$_3$ in terms of the  amplitude of the U parameter. 
Atomic distortions are quantified from a symmetry-adapted mode analysis (see Table S1). The $R_3^-$ and 
$M_5^+$ distortions which have negligible amplitudes (< 0.05 $\AA$) are not shown.}
  \label{figsupp:struct1}
\end{figure}

\begin{table}[h!]
\begin{tabular}{lccc}
      \hline
      \hline
      {Labels} &{Atomic motion} & ${Pbnm}$ & ${P2_1/n}$\\
      \hline
     $R_5^-$ &{\bf Anti-phase rotations of O octahedra along $x$ and $y$ ($R_{xy}$)} & {\bf x} & {\bf x} \\
     $M_2^+$ & {\bf In-phase rotations of O octahedra along $z$ ($M_z$)} &{ \bf x} & {\bf x} \\
     $X_5^-$ &Anti-polar (layered) motion of $R$ cations ($X_{AP}$)  & {x} & {x} \\
     $M_3^+$ &Jahn-Teller distortion of O octahedra ($Q_2^+$) & x & x \\
     $R_4^-$ &Anti-polar motion (rocksalt) of $R$ cations ($R_{AP}$) &x & x \\
     $R_2^-$ & {\bf Breathing distortion of the O octahedra ($B_{OC}$)} &   & {\bf x} \\
     $M_5^+$ &Anti-polar motion of O ($M_{AP}$) &  & x \\
     $R_3^-$ &Jahn-Teller distortion of O octahedra ($Q_2^-$) &  & x \\
      \hline
      \hline
  \end{tabular}
  \caption{Labels and description of the distortions of the $Pm\bar{3}m$ phase allowed by symmetry in the $Pbnm$ and $P2_1/n$ 
  phases of $R$NiO$_3$ compounds. The main distortions are in bold.}
\end{table}

We see in Figure S1 that the lattice parameters are rather independent of U and well described within the whole range (error smaller
than $1\%$).   At the level of the atomic distortions, the amplitude of breathing mode $B_{OC}$ ($R_2^-$) is only properly described in 
the limit  of small U values.  For the dominant modes like $R_{xy}$ ($R_5^-$) or $M_z$ ($M_2^+$), although the relative errors remain 
reasonably small for any U,  the absolute amplitude evolves significantly with U and also converge to the correct values at low U.
  
\begin{figure}
 \includegraphics[width=0.8\linewidth]{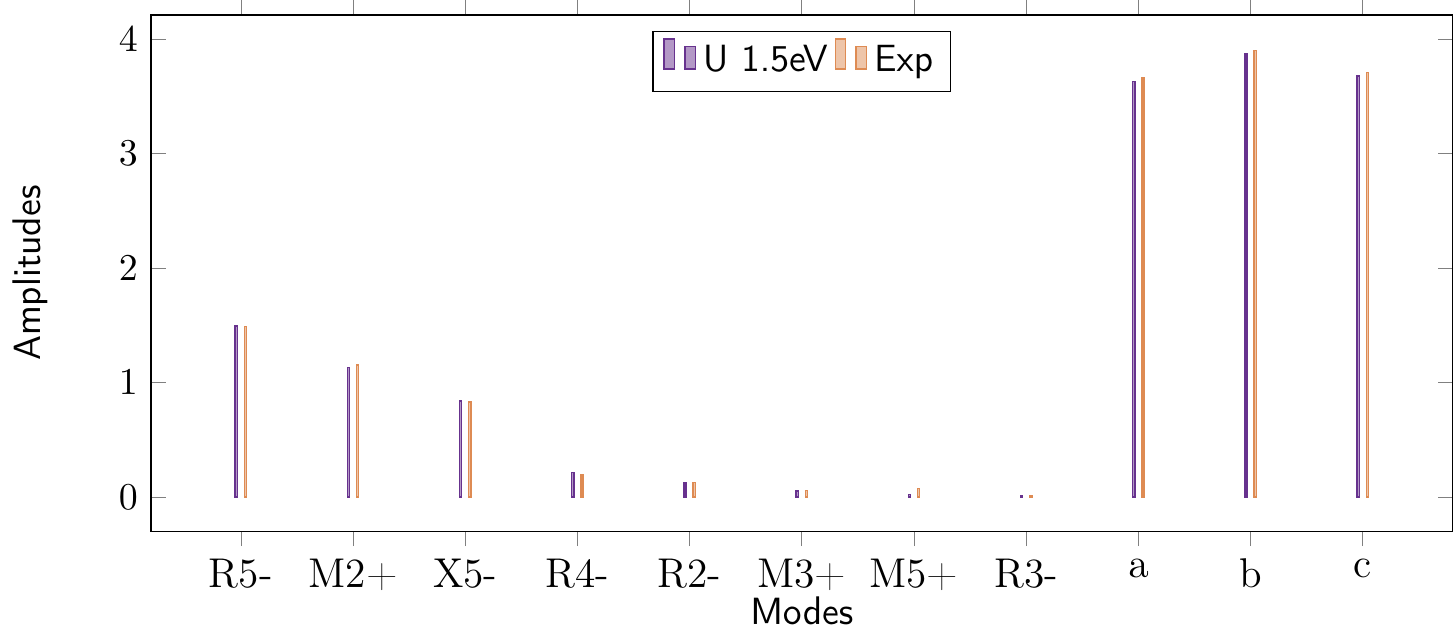}
 \caption{Comparison of the absolute amplitudes (\AA)  of the atomic distortions and lattice parameters in the E'-type AFM $P2_1/n$ phase of YNiO$_3$ 
 as computed in DFT with U = 1.5 eV  (purple) and as measured experimentally (orange). The E'-type AFM $P2_1/n$ ground state has a 80-atoms unit cell 
 with lattice parameters $(a',b',c') = (2 \sqrt{2} a, \sqrt{2} b, 4c)$. }
 \label{figsupp:struct2}
\end{figure}
  
In Figure S2, we report comparison with experiment data \cite{Alonso2001}  of the absolute amplitudes of the atomic distortions and lattice parameters 
in the E'-type AFM $P2_1/n$ phase of YNiO$_3$ as computed in DFT with U = 1.5 eV. It confirms that the atomic structure of YNiO$_3$ is very 
accurately described in DFT using PBESol and a U parameter of 1.5 eV.

\subsection{\label{subsec3:levelsupp1}B. Magnetic properties}

\begin{figure}
 \includegraphics[width=0.8\linewidth]{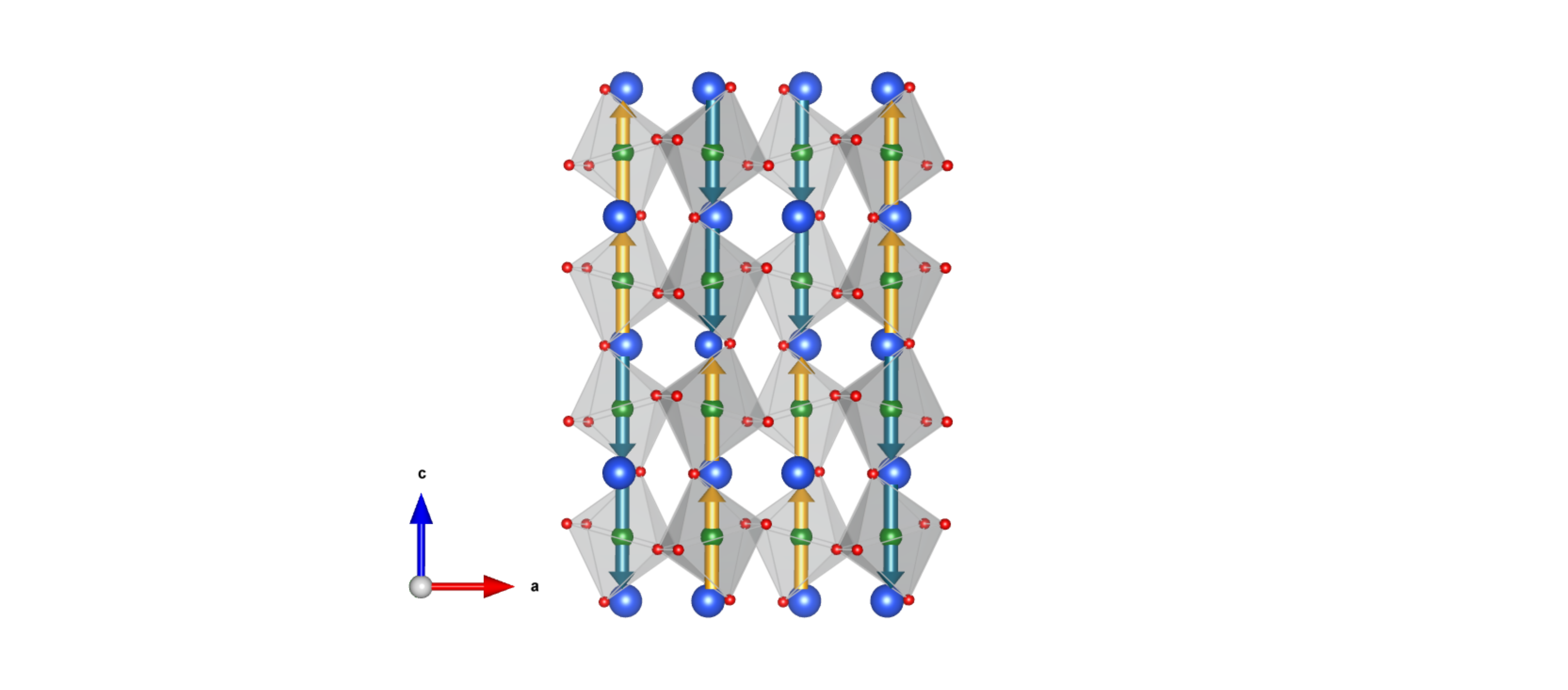}
 \caption{Spin order associated to the E'-type antiferromagnetic (AFM-E') ground state.}
 \label{figsupp:magn1}
\end{figure}

In order to determine the magnetic ground state of the $P2_1/n$ phase of YNiO$_3$, we performed calculations for various magnetic orders associated to supercells of
up to 80 atoms. While U values larger than 2 systematically favour a FM spin order, U = 1.5 eV properly stabilizes the E'-type spin ordering illustrated in Figure S3 as the
ground state. It corresponds to an ``up-up-down-down'' spin arrangement related to a Bragg vector ${\bf q} = (1/4,1/4,1/4)$ in pseudocubic notations. 

In our calculations, we get a magnetic moment $\mu = 1.2 \mu_B$ on the Ni atoms associated to the large oxygen octahedra and $\mu \approx 0 \mu_B$ on the 
Ni atoms associated to the small octahedra. This is similar with what has been reported in Ref. \cite{Prosandeev2012,Varignon2016} and in line with the $d^8-d^8L^2$ picture 
\cite{Johnston2014}. It is also compatible with experimental data as discussed in Ref. \cite{Alonso2009}.

\begin{figure}
 \includegraphics[width=0.6\linewidth]{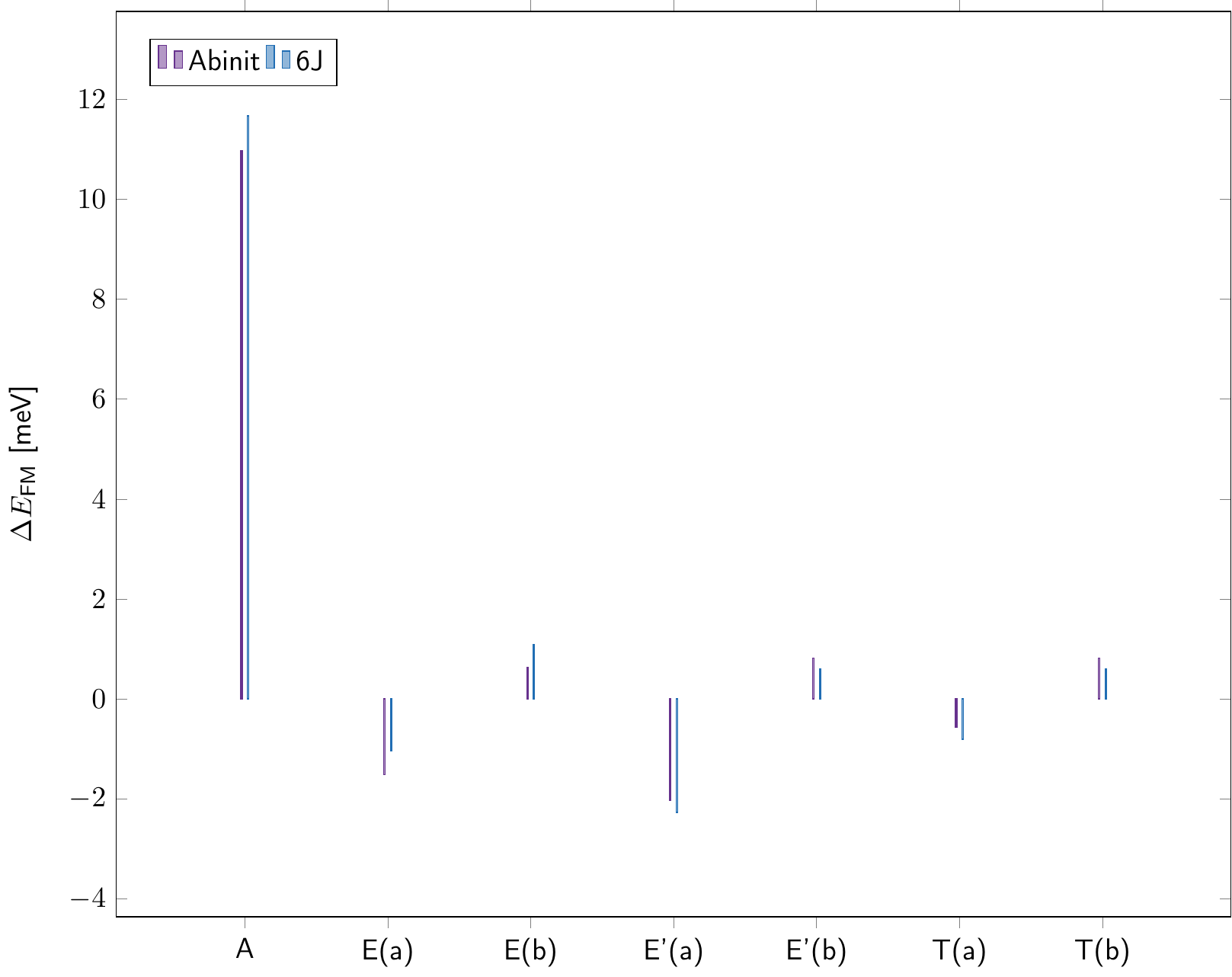}
 \caption{Comparison of the energy associated to various spin orders, respect to the FM order taken as reference, as obtained from our DFT calculations (purple) and simple spin model (blue). The A-type ordering corresponds to an antiferromagnetic stacking along the $c$ direction of FM ordered $ab$ planes. The E(a)-type order corresponds to an ``up-up-down-down'' zigzag chain along the $a$ (orthorhombic) direction, and a FM order along the $c$ direction. The E(b)-type order is characterised by the same ``up-up-down-down'' zigzag chain but along the $b$ (orthorhombic) direction. The E'(a), E'(b), T(a) and T(b) types of order have the same in-plane properties too. However, the stacking along the $c$ direction is different : E'(a) and E'(b) have two FM-order layers,  AFM coupled with the next two layers while in the T(a) and T(b) configurations, the zigzag chain shifts about one spin along the chain direction at the same time one passes throught the $c$ direction.  The ground state is E'(a) as illustrated in Figure S3 and simply called E' within the manuscript.} 
  \label{figsupp:mag}
\end{figure}

Beyond the fact that DFT calculations with U = 1.5 eV provides the right magnetic ground state, it is interesting to check if it properly accounts for the 
strength of the magnetic interactions. To that end, we built a simple spin model $E_{mag}= (1/2) \sum J_{ij} S_i S_j$  including $J_{ij}$ interactions 
up to fourth neighbours (6 independent parameters) and fitted the parameters on our first-principles data \cite{Mercy2017}. As illustrated in Figure S4 for YNiO$_3$
this spin model properly reproduces the energetics of the first-principles calculations.

Monte-Carlo simulations (using large boxes up to 1728 Ni atoms) from this spin-model~\cite{Mercy2017} (i) confirmed the E'-type ground state and (ii)  provided a Neel 
temperature $T_N =$ 154 K, very similar to the mean-field estimate of 166 K and in close agreement with the experimental value of 150 K for YNiO$_3$ \cite{Alonso1999}.  

This demonstrates that our DFT calculations with U = 1.5 eV reproduces not only the correct E'-type magnetic ground state of nickelates but also properly
describes the strength and anisotropy of their magnetic interactions.

\subsection{\label{subsec2:levelsupp1}C. Electronic properties}

Our DFT calculations with U = 1.5 eV properly accounts for the insulating character of the E'-type AFM $P2_1/n$ ground state of YNiO$_3$.  
For the electronic bandgap, we get a value of 0.46 eV in reasonable agreement with the experimental estimate of 0.305 eV \cite{Arima1993}. 

The electronic properties are further discussed in the manuscript.  As it appears clearer there, the structural and electronic properties are intimately linked 
together in nickelates. Hence, the fact that our simulations describe accurately the structural properties of these compounds strongly suggests that they can also be trusted to investigate their electronic properties.

\section*{\label{sec:levelsupp1prime}S2. Phonon dispersion curves}

\begin{figure}
  \includegraphics[width=0.7\linewidth]{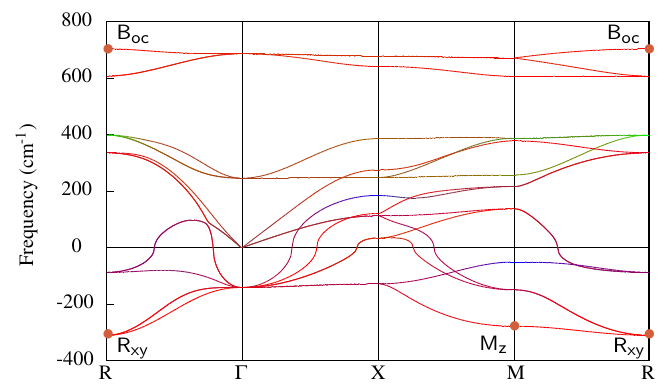}
 \caption{Phonon dispersion curve for the $Pm\bar{3}m$ phase of YNiO$_3$, at the $P2_1/n$ ground state volume and for a FM spin ordering (U=1.5 eV).
 Colors have been associated to the curves according to the involvement of each cation in the eigendisplacement of each mode ($R$ in blue, Ni in green and O in red).
 Imaginary frequencies (unstable modes) appear as negative values.}
 \label{phonon}
\end{figure}

In Figure S5, we report the full phonon dispersion curves of the $Pm\bar{3}m$ phase of YNiO$_3$, as calculated for a FM spin ordering at the volume of the $P2_1/n$ 
AFM-E' phase ($a_{pc} = 3.728$\AA). Similar curves have been obtained at the relaxed volume ($a_0 = 3.695$\AA). Interpolation of these phonon dispersion curves 
relies on the calculation of the interatomic force constants within a $2\times2\times2$ supercell. Although this might not be totally sufficient to get a fully converged 
interpolation, it provides already a good estimate of the shape of the dispersion curves. It is worth to notice that the frequencies at the high-symmetry points, which are the 
only ones discussed below and in the manuscript, are not interpolated but calculated explicitly within our approach.  

On the one hand, the phonon dispersion curves highlight strong instabilities at the R and M $q-$points of the BZ, associated to the $R_{xy}$ and $M_z$ distortions. On the other hand, and quite amazingly, the breathing distortion, $B_{OC}$ located at the R point, which finally produces the MIT, is associated to the hardest mode in the cubic phase. Clearly, such hard mode  cannot spontaneously condense within the cubic structure.

We further notice the presence of two (triply-degenerated) unstable modes at $\Gamma$ with very close frequencies. The softest one is associated to a polar-type motion involving 
$R$ and O atoms; this kind of instability is expected for perovskites with small tolerance factors as YNiO$_3$. The second one is the silent butterfly motion of the O atoms.

\section*{\label{sec:levelsupp2}S3. Landau Model}

Based on our DFT results, we have built a simple Landau-type model. In order to highlight the triggered mechanism, we restricted ourselves to the simplest possible model  including only 
$R_{xy}$, $M_z$ and $B_{OC}$ degrees of freedom and lowest-order terms. 

\subsection{\label{subsec1:levelsupp2}Expression}
 
Within our Landau-type model, the free energy in terms of the amplitudes $Q_R$, $Q_M$ and $Q_B$ (of $R_{xy}$, $M_z$ and $B_{OC}$ respectively) reads:

\begin{eqnarray}
E(Q_M,Q_R,Q_B)=& \alpha_R Q_R^2 + \beta_R Q_R^4 +  \alpha_M Q_M^2 + \beta_M Q_M^4 +  \alpha_B Q_B^2 + \beta_B Q_B^4 \notag\\ &+ \lambda_{MR} Q_M^2 Q_R^2 + \lambda_{MB} Q_M^2 Q_B^2 + \lambda_{RB} Q_R^2 Q_B^2 
\end{eqnarray}
The parameters $\alpha_R$ and $\alpha_M$ are assumed to be temperature dependent as 
\begin{eqnarray}
\alpha_R = \gamma_R (T_{0R}-T) \;\;\; {\rm and} \;\;\;   \alpha_M = \gamma_M (T_{0M}-T) 
\end{eqnarray}
while all the other parameters are supposed to be constant.

Other modes allowed by symmetry (see Table S1) in the $Pbnm$ and $P2_1/n$ phases have not be explicitly included within the model. Some of them, like $X_5^-$ (and to a lesser extent $R_4^-$ and $M_3^+$), take however a significant amplitude and are crucial to stabilize the $Pbnm$ phase. They are implicitly included through a renormalization of the $\lambda_{MR}$ parameter as it will appear more clearly in the next subsection.

The strain degrees of freedom have not been explicitly included within the model to highlight the key role of phonon-phonon couplings, which appear sufficient to reproduce experimental data on bulk compounds. However, this model could be naturally extended to strain degrees of freedom and their couplings with lattice modes. This might be useful to quantify for instance the role of epitaxial strain in thin films but is beyond the scope of this work.   

The expansion has been limited to 4$^{th}$ order for all three order parameters including $Q_B$. This is justified by the fact that, from the fit of the parameters, the triggered transition appears to be second order . We notice however that explicit treatment of the strain (neglected here) could affect the order of the  phase transition as further discussed below.

\subsection{\label{subsec2:levelsupp2}Fit from DFT}

Parameters of our Landau-type model have been fitted on first-principles results.  At first, we focused on YNiO$_3$. 

\begin{figure}
  \includegraphics[width=0.5\linewidth]{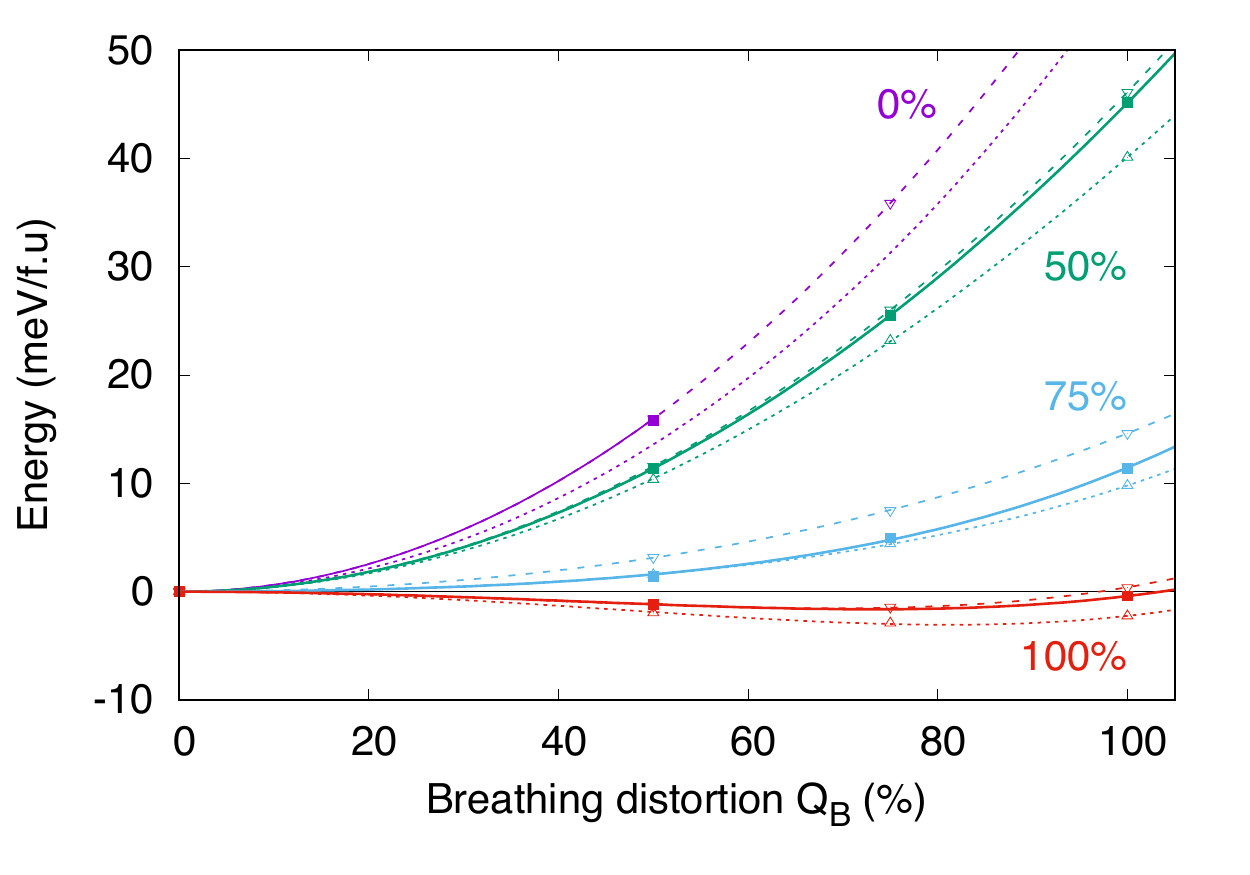}
 \caption{Evolution of the energy of YNiO$_3$ in terms of the amplitude of the breathing distortion for fixed amplitude of oxygen rotations ($Q_R=Q_M$, from 0\% to 100\%) in a FM configuration and either a fixed cubic cell (dotted line and triangles: $a_{pc} = 3.728$ \AA; dashed line and back triangles: $a_0 = 3.695$ \AA) or when relaxing fully the lattice parameters (full line and squares).}
 \label{figure-SR}
\end{figure}

We considered in our calculations a fixed cubic $Pm\bar{3}m$ cell at a volume similar to that of the $P2_1/n$ AFM-E' ground-state ($a_{pc}= 3.728$\AA), which corresponds to imposing a negative strain of $0.9 \%$. At this volume, relaxing within the $P2_1/n$ symmetry while keeping the unit cell fixed yields amplitudes of distortion comparable to the ground-state. We notice that, as illustrated in Fig. \ref{figure-SR}, similar calculations performed at the relaxed lattice constant ($a_0 = 3.695$\AA) yield very similar results. Even calculations performed while relaxing the lattice parameters at fixed mode amplitudes (in reduced coordinates) do not provide any significant change. 

The calculations have been performed with a FM spin order which does not break any symmetry. We checked explicitly that the key physical features and conclusions (cooperative bi-quadratic coupling between rotations and breathing and triggered mechanism) remain similar for different AFM spin orders. The results remain even very similar in a non-magnetic (NM) calculation (with or without U correction) although, in that case, the amplitude of rotations required to destabilise $B_{OC}$ is slightly larger ($\approx 160$\%); this last result illustrates that electronic Hund's rule energy, although playing a role, is not driving alone the appearance of $B_{OC}$ as sometimes suggested \cite{Mazin2007}.

The parameters of the Landau model at 0 K have then been extracted from DFT data as follows.
\begin{itemize}
  \item $\alpha_R^0 (=\gamma_{R} T_{0R}$), $\alpha_M^0 (=\gamma_{M} T_{0M}$), $\beta_R$ and $\beta_M$ were fitted on the individual double wells associated to  $R_{xy}$ and $M_z$ (Figure 1).
  \item $\lambda_{RM}$ was fitted to reproduce the energy of a relaxed $Pbnm$-like phase (full atomic relaxation while keeping the cubic cell fixed). From this, we renormalize the natural competition between $R_{xy}$ and $M_z$ by including implicitly the stabilising effect of $X_5^-$, $R_4^-$ and $M_3^+$ modes. We notice that in all compounds, $R_{xy}$ and $M_z$ compete with each other and should yield $\lambda_{RM}>0$. However, because of the renormalization due to the implicit presence of the other modes, $\lambda_{RM}$ becomes negative for large cations (i.e. $X_5^-$ helps stabilizing the $Pbnm$ phase consistently with the discussion in Ref. \cite{Benedek2013}.
  \item $\alpha_B$ was fitted on the single well associated to  $B_{OC}$ (Figure 1). 
  \item $\lambda_{BR}$ and $\lambda_{BM}$ were fitted from the change of curvature of the well of $B_{OC}$ when freezing 100\% of $Q_R$ and $Q_M$ respectively (Figure 2).
  \item $\beta_B$ was fitted to reproduce the right amplitude of $B_{OC}$ in the ground state of the model and it was checked that the result still properly describes the single well associated to  $B_{OC}$.
\end{itemize}

Within the model, the amplitude for the atomic distortion are renormalised to the one obtained from DFT calculation for the YNiO$_3$ ground state.
This means 1 for rotation, tilts and breathing mode correspond to the amplitude of these modes in a cubic box with lattice parameters coresponding to 3.728 \AA.

We applied the same procedure to GdNiO$_3$ and SmNiO$_3$. All the computed parameters are summarized in Table \ref{fit}.

\begin{table}
  \begin{tabular}{cccc}
\hline
\hline
    parameter & Y & Gd & Sm \\
\hline
$t$ & 0.920 &0.938  &0.947 \\
\hline
$\alpha_B$&58.1&52.6&50.6\\
    $\beta_B$&10.0 &16.0 &31.0 \\
    $\alpha_M^0$& -578.3& -385.5& -277.3\\
    $T_{0M}$&3918 &2571 &1897 \\
    $\beta_M$&213.9 & 209.4 & 201.0 \\
    $\alpha_R^0$ &-1288.1 &-1081.2&-920.9\\
    $T_{0R}$&3918 &3195 &2833 \\
    $\beta_R$ &   648.8&754.9& 750.4\\
    $\lambda_{MR}$ & 31.8 & -57.7&-99.4\\
    $\lambda_{MB}$ &-26.0  &-29.0  & -29.4\\
    $\lambda_{RB}$ &-35.8 & -42.2 & -42.4 \\
\hline
\hline
  \end{tabular}
  \caption{Landau model parameters (meV/f.u. or K) as fitted on first-principles data, using the mode normalisation described in the Methods section.}
  \label{fit}
\end{table}

\begin{figure}
  \includegraphics[width=1.0\linewidth]{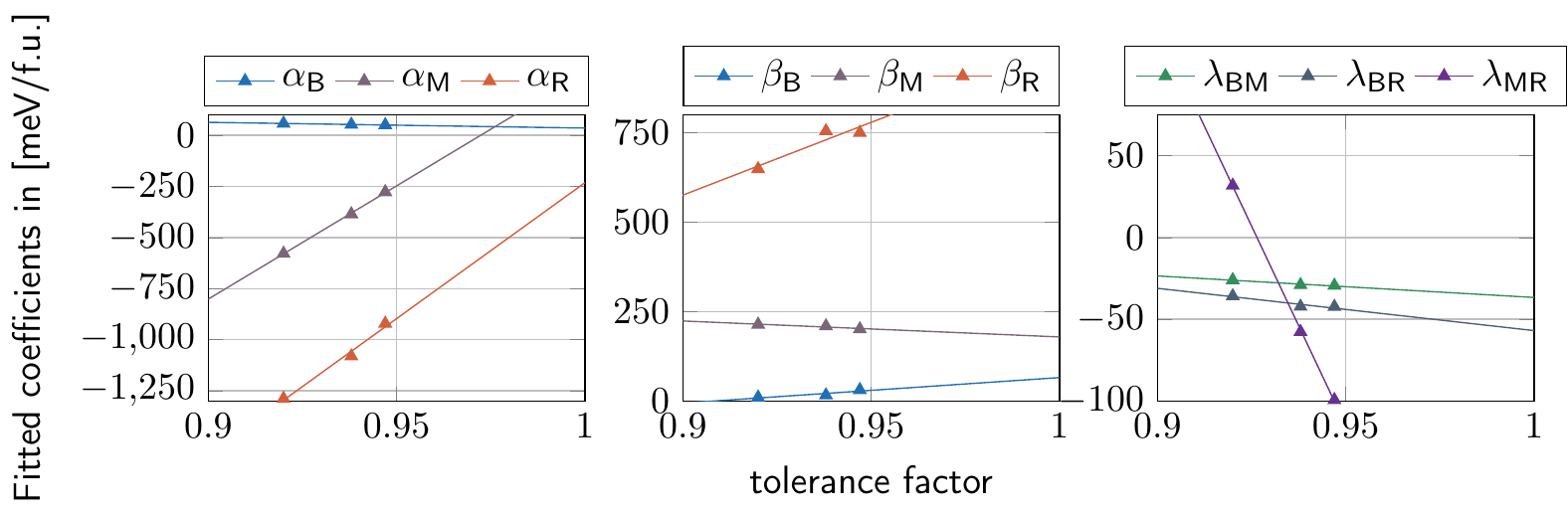}
 \caption{Evolution of the model parameters (meV/f.u.) with the tolerance factor.}
 \label{param}
\end{figure}


As illustrated in Figure S6, all the parameters have an almost linear dependence in terms of the tolerance factor $t$. 
So, in our model, we assumed such a linear dependence to determine the value of the parameters at arbitrary $t$.

Finally, knowing $\alpha_{R,M}$ at 0 K (from the DFT calculations), their temperature dependence was estimated as follows. 
Focusing first on YNiO$_3$ we adjusted $T_{0R}=T_{0M}$ so that within our model, $B_{OC}$ appears at the 
experimental value of 585K \cite{Catalan2008} and we deduced $\gamma_{M,R} = \alpha^0_{R,M}/T_{0R,M}$:  $\gamma_{M} = 0.148 meV/f.u.$
and $\gamma_{R} = 0.331 meV/f.u.$. Then, assuming 
$\gamma_{M,R}$ constant within the whole family the evolution of $T_{0R}$ and $T_{0M}$ with the tolerance factor were 
obtained as:  $T_{0R,M} = \alpha^0_{R,M}/\gamma_{M,R}$. 

From this, we get the following final Landau-type expression  for the free energy, allowing us to determine the values of 
$Q_R$,$Q_M$ and $Q_B$ for any value of the temperature $T$ and tolerance factor $t$ :

\begin{eqnarray}
  &E(Q_M,Q_R,Q_B)=\notag \\& (-281.9\times t+317.4) Q_B^2  + (714.3\times t-648.9) Q_B^4 \notag\\& -0.15 \times  (-74891.2\times t+72818.6 - T)\times  Q_M^2 + (-446.3\times t+625.3)\times _M^4  \\& -0.33 \times   ( -40212.2\times t+40913.9  - T)\times Q_R^2 +( 4067.9\times t-3085.5)\times Q_R^4 \notag\\& +(-4875.7\times t+4517) Q_M^2 Q_R^2  +(-131.6\times t+94.9) \times Q_M^2 Q_B^2 +(-258.6\times t+201.6)\times   Q_R^2 Q_B^2 \notag
\end{eqnarray}

\subsection{\label{subsec1:levelsupp2}Analytical solution}

From this model, $T_{MI}$ can be determined analytically. 

We start from the expression:

\begin{eqnarray}
  E(Q_M,Q_R,Q_B)=& \alpha_B Q_B^2 + \beta_B Q_B^4 \notag\\&+ \gamma_M \times (T_{0M} - T)\times  Q_M^2 + \beta_M Q_M^4  \notag\\& +\gamma_R \times (T_{0R} - T)\times Q_R^2 + \beta_R Q_R^4 \notag\\&+ \lambda_{MR} Q_M^2 Q_R^2 + \lambda_{MB} Q_M^2 Q_B^2 + \lambda_{RB} Q_R^2 Q_B^2 
\end{eqnarray}

At the energy minimum, we should have :

\begin{equation}
    \frac{\partial E}{\partial Q_M}=0 \, \, {\rm and} \,\, \frac{\partial E}{\partial Q_R}=0.
\end{equation}

The solutions for that, other than $Q_M=0$ and $Q_R=0$, are :

\begin{eqnarray}
  Q_M^2=& \frac{-2 T_{0M} \gamma_M \beta_R + T_{0R} \gamma_R \lambda_{MR} + T (2 \gamma_M \beta_R - \gamma_R \lambda_{MR}) }{4 \beta_M \beta_R - \lambda_{MR}^2}\\ \notag +&\frac{  (-2 \beta_R \lambda_{MB} + \lambda_{MR} \lambda_{RB})}{4 \beta_M \beta_R - \lambda_{MR}^2} Q_B^2\\\notag
  Q_R^2=& \frac{-2 T_{0M} \gamma_M \beta_R + T_{0R} \gamma_R \lambda_{MR} + T (2 \gamma_M \beta_R - \gamma_R \lambda_{MR})}{4 \beta_M \beta_R - \lambda_{MR}^2}\\ \notag +&\frac{(-2 \beta_R \lambda_{MB} + \lambda_{MR} \lambda_{RB})}{ 4 \beta_M \beta_R - \lambda_{MR}^2} Q_B^2
\end{eqnarray}

Introducing this in Eq. (4) we get :

\begin{equation}
  E(Q_B)= \alpha_B^{\prime} Q_B^2 + \beta_B^{\prime} Q_B^4
\end{equation}

where

\begin{eqnarray}
  \alpha_B^{\prime}=& \alpha_B  + \frac{(\gamma_M \lambda_{MR} \lambda_{RB}-2\gamma_M \beta_R \lambda_{MB}) }{ 4 \beta_M \beta_R - \lambda_{MR}^2}  (T_{0M}-T)\\ \notag+& \frac{ (\gamma_R \lambda_{MB} \lambda_{MR} - 2 \gamma_R \beta_M \lambda_{RB})}{ 4 \beta_M \beta_R - \lambda_{MR}^2} (T_{0R}-T) \\
  \beta_B^{\prime}=& \beta_B +  \frac{ \lambda_{MB} \lambda_{MR} \lambda_{RB} - \beta_M \lambda_{RB}^2 - \beta_R \lambda_{MB}^2}{4 \beta_M \beta_R - \lambda_{MR}^2} \\ 
  \label{betaprime}
\end{eqnarray}

The MIT is linked to the appearance of the $B_{OC}$. This will appear at a temperature $T_{MI}$ at which $\alpha_B^{\prime}=0$.  
This critical temperature is given by :

\begin{eqnarray}
  T_{MI}=&\frac{ \alpha_B (-4 \beta_M \beta_R + \lambda_{MR}^2)}{2 \gamma_M \beta_R \lambda_{MB} - \gamma_R \lambda_{MB} \lambda_{MR} +  2 \gamma_R \beta_M \lambda_{RB} - \gamma_M \lambda_{MR} \lambda_{RB}}\\ \notag  +& \frac{  T_{0R} \gamma_R (-\lambda_{MB} \lambda_{MR} + 2 \beta_M \lambda_{RB})}{2 \gamma_M \beta_R \lambda_{MB} - \gamma_R \lambda_{MB} \lambda_{MR} +  2 \gamma_R \beta_M \lambda_{RB} - \gamma_M \lambda_{MR} \lambda_{RB}} \\ \notag+& \frac{T_{0M} \gamma_M (2 \beta_R \lambda_{MB} - \lambda_{MR} \lambda_{RB}))}{2 \gamma_M \beta_R \lambda_{MB} - \gamma_R \lambda_{MB} \lambda_{MR} +  2 \gamma_R \beta_M \lambda_{RB} - \gamma_M \lambda_{MR} \lambda_{RB}}
\end{eqnarray}

Furthermore, supposing a linear dependence for all the coefficients with respect to the tolerance factor, we get a generic expression :

\begin{equation}
  T_{MI} =   \frac{a+ t\times (b+(c+d\times t)t)}{e+t\times (f+g\times t)}
\end{equation}
where $t$ is the tolerance factor and $a,b,c,d,e,f$ and $g$ are a combination of model parameters.

Using the coefficients determined from DFT calculations in the previous Section, we can predict the evolution of $T_{MI}$ as a function of the tolerance factor as illusrated in 
Figure 2b, blue line.

Independently, we can also fit the experimental data point using Eq. 13. Making such a fit, while excluding Nd and Pr compounds, we get the dashed blue line in Figure 2b.

\subsection{\label{subsec2:levelsupp2}Order of the transition}

Experimentally, there is still some debate about the order of the MIT. For large cations ($T_{MI}=T_N$), the MIT is rather abrupt and hysteretic and unanimously considered as being first order \cite{Catalan2008}. The magnetic transition that takes place at the same temperature is also first order \cite{Vobornik1999}. For small cations ($T_{MI}>T_N$), the MIT is less hysteretic and sometimes considered as evolving to second-order. Some studies seem however to show that it stays first-order \cite{Perez1999,Nikulina2004}, while the less hysteretic behavior could be related to the fact that kinetics are better at higher temperatures \cite{Catalan2008}. For these compounds the magnetic transition is second-order.

As previously mentioned, the MIT is predicted to be second-order within our very simple model. As highlighted in Table S2, computed $\beta_B$ is positive for all compounds. From Eq. \ref{betaprime}, the oxygen rotations renormalize  the fourth-order term coefficient and a negative value of $\beta'_B$ would give rise to a first-order transition (it would then further require including 6$^{th}$ order terms in $Q_B$). Although this renormalization is negative, $\beta'_B$ stays nevertheless positive in all cases ($\beta'_B =$ 9, 14 and 29 respectively for YNiO$_3$, GdNiO$_3$ and SmNiO$_3$ respectively) corresponding therefore to a second-order transition.

Yet, we have to stress that our approach does not allow us to address the order of the transition conclusively. First, our model is built at fixed cubic cell and does not include strain relaxation. Explicit treatment of the latter will further renormalize the $4^{th}$-order term and might potentially make it negative, so eventually changing the order of the transition. Second, at a more fundamental level, even if our DFT+U results suggest that $Q_B$ undergoes a second-order transition, that does not rule out the possibility that thermal effects effectively render a first-order transformation driven by temperature. The ferroelectric phase transitions of BaTiO$_3$, a well studied case, are a concrete example of this \cite{Iniguez2001}, and also illustrate the critical role of strains to enhance the discontinuous character of the transformation \cite{Zhong1995}. Hence, discussing the character of the transition from first-principles would require explicitly statistical simulations that fall beyond of the scope of this work. The Landau model introduced here  was kept simple on purpose in order to highlight the key role of the triggered mechanism and it show that, based on our first-principles results and  a minimal experimental input, the main features of the phase diagram can be readily reproduced. Further, the main trends (regarding ionic size, amplitude of the different rotations, ...) are properly captured by this simple model, which moreover provides us with insights about how to tune the behaviour of these materials. In this sense, we consider that the proposed Landau model is valid and useful although, admittedly, it is not suitable for a definite discussion of the order of the transition.

For large cations (Nd and Pr), our model predicts that the MIT can no more be fully triggered by the oxygen rotations (which are reduced); for those compounds, it is complementarily promoted by the appearance of the AFM-E' magnetic order. In such case, the MIT takes place at $T_{MI}= T_N$; it is expected to be more abrupt and to be first order. Such coupling between the structural and magnetic transition for large cations is in line with the conclusions of Vobornik {\it et al.} in Ref. \cite{Vobornik1999}: they suggest indeed that, contrary to other cases,  there is a possible interplay between electronic and magnetic degrees of freedom when $T_{MI}= T_N$ and that any further model of the TMI should address that fact. Our manuscript explicitly addresses that point and we believe that it convincingly answers their questioning.

\section*{\label{sec:levelsupp4}S4. Electronic band structures}

In Figure S6, we report the electronic dispersion curves of YNiO$_3$ with a FM spin order, along a more exhaustive path of the Brillouin zone of the $Pbnm$ or $P2_1/n$ 20-atom cell.  The majority spins are in colors while the minority spins are in light grey. The latter have been omitted for clarity in the main manuscript. We notice that the cubic phase is essentially non magnetic (up and down spin bands nearly degenerate) and magnetism starts to develop with the rotations.  

In Figure S7, we report similarly the electronic dispersion curves of YNiO$_3$ but with an AFM-A spin order. This figure is very similar to the previous one, demonstrating that our results are not dependent of the specific choice of spin order.

\begin{figure}
  \includegraphics[width=0.6\linewidth]{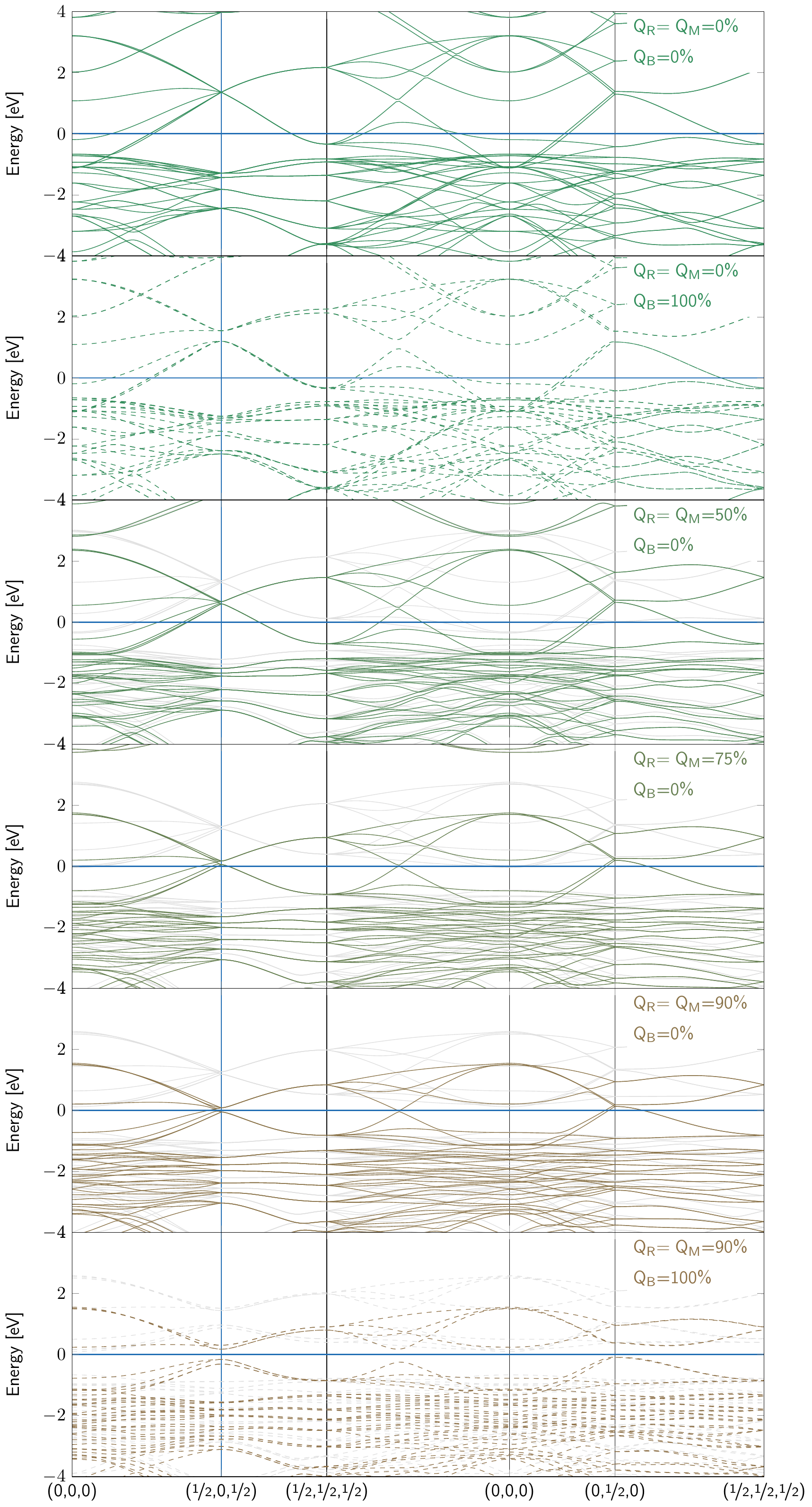}
  \caption{YNiO$_3$, FM spin order. Evolution of the electronic dispersion curves around the Fermi energy $E_f$  when 
  freezing into the $Pm\bar{3}m$ phase increasing amplitudes of oxygen rotations  ($Q_R = Q_M$ from 0\% to 90\%, lines) and eventually adding the breathing distortion ($Q_B = 100$\%, dashed lines). The graph connects high-symmetry points in the Brillouin zone of the $Pbnm$ or $P2_1/n$ 20-atom cell. Majority spins are in colors and minority spins in light grey. }
  \label{figsupp:breathingwell}
\end{figure}

\begin{figure}
  \includegraphics[width=0.6\linewidth]{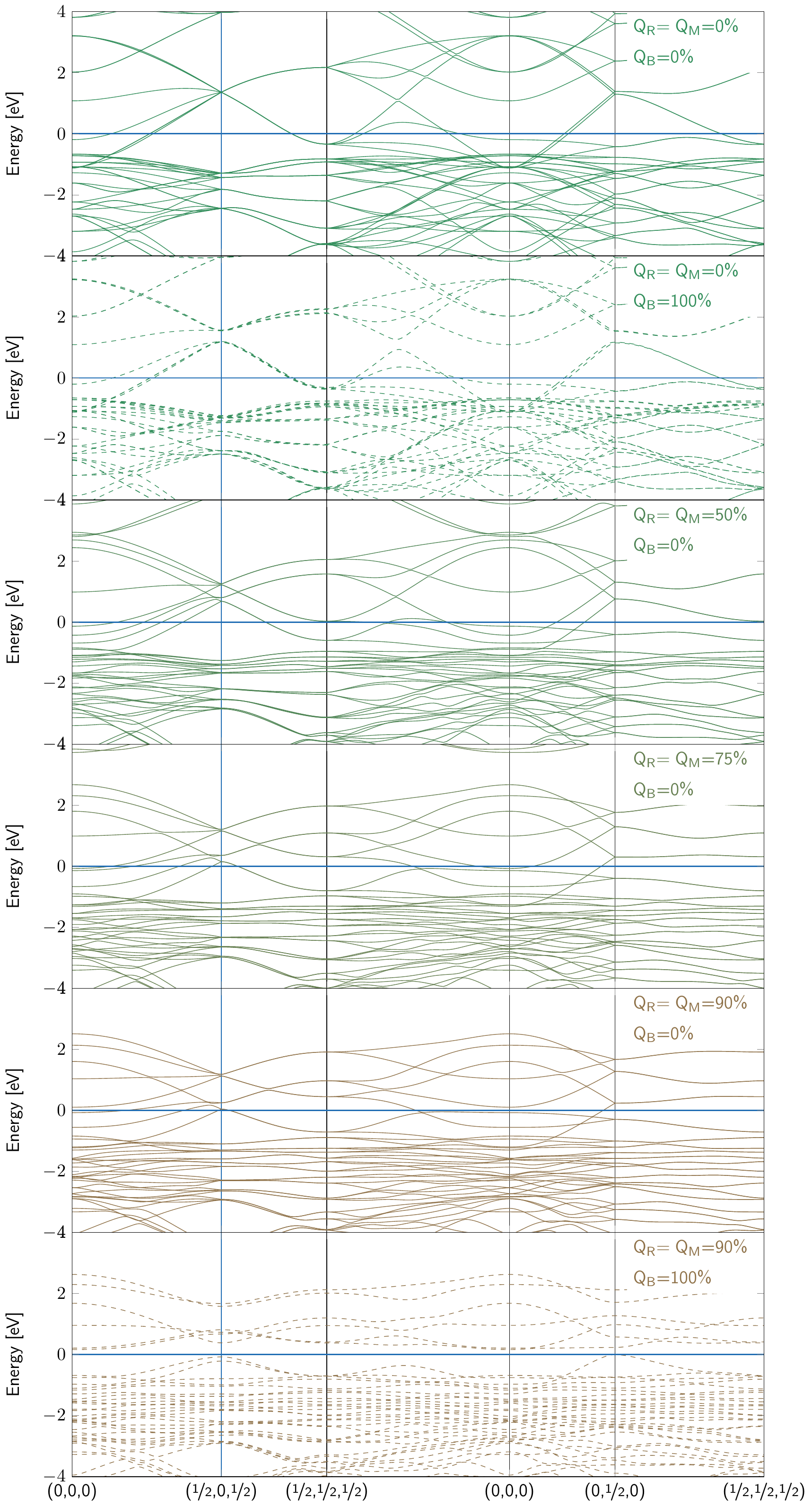}
  \caption{YNiO$_3$, AFM-A spin order. Evolution of the electronic dispersion curves around the Fermi energy $E_f$  when 
  freezing into the $Pm\bar{3}m$ phase increasing amplitudes of oxygen rotations  ($Q_R = Q_M$ from 0\% to 90\%, lines) and eventually adding the breathing distortion ($Q_B = 100$\%, dashed lines). The graph connects high-symmetry points in the Brillouin zone of the $Pbnm$ or $P2_1/n$ 20-atom cell.}
  \label{figsupp:breathingwell}
\end{figure}

\newpage

\end{document}